\documentclass[prd,aps,a4,twocolumn,superscriptaddress,preprintnumbers,nofootinbib]{revtex4-1}
\usepackage{pslatex}
\usepackage[pdftex]{graphicx}
\usepackage{psfrag}
\usepackage{epsfig}
\usepackage{color}
\usepackage{cancel}
\usepackage{slashed}
\usepackage{amssymb}
\usepackage{amsmath}
\usepackage{hyperref}
\usepackage{enumerate}
\usepackage{multirow}
\newcommand{\cvn}{CE$\nu$NS}
\newcommand{\cosr}{$\cos \theta_r$}

\begin{document}
\begin{center}
\hfill MI-TH-209
\end{center}
\title{Coherent Elastic Neutrino-Nucleus Scattering with directional detectors}
\author{M. Abdullah}
\email{mabdullah@tamu.edu}
\affiliation{Department of Physics and Astronomy, Mitchell Institute
  for Fundamental Physics and Astronomy, Texas A\&M University,
  College Station, TX 77843, USA}
\author{D. Aristizabal Sierra}%
\email{daristizabal@ulg.ac.be}%
\affiliation{Universidad T\'ecnica
  Federico Santa Mar\'{i}a - Departamento de F\'{i}sica\\
  Casilla 110-V, Avda. Espa\~na 1680, Valpara\'{i}so, Chile}%
\affiliation{IFPA, Dep. AGO, Universit\'e de Li\`ege, Bat B5, Sart
  Tilman B-4000 Li\`ege 1, Belgium}%
\author{Bhaskar Dutta}%
\email{dutta@physics.tamu.edu}%
\affiliation{Department of Physics and Astronomy, Mitchell Institute
  for Fundamental Physics and Astronomy, Texas A\&M University,
  College Station, TX 77843, USA}%
\author{Louis E. Strigari}%
\email{strigari@tamu.edu}%
\affiliation{Department of Physics and Astronomy, Mitchell Institute
  for Fundamental Physics and Astronomy, Texas A\&M University,
  College Station, TX 77843, USA}%
\begin{abstract}
  We study the sensitivity of detectors with directional sensitivity
  to coherent elastic neutrino-nucleus scattering (CE$\nu$NS), and how
  these detectors complement measurements of the nuclear recoil
  energy. We consider stopped pion and reactor neutrino sources, and
  use gaseous helium and fluorine as examples of detector material. We
  generate Standard Model predictions, and compare to scenarios that
  include new, light vector or scalar mediators. We show that
  directional detectors can provide valuable additional information in discerning
  new physics, and we identify prominent spectral features in both the 
  angular and the recoil energy spectrum for light mediators, even for 
  nuclear recoil energy thresholds as high as $\sim 50$ keV. Combined with energy 
  and timing information, directional
  information can play an important role in extracting new physics
  from CE$\nu$NS experiments.
\end{abstract}
\maketitle
% ----------------
% Sec: motivation
% ----------------
\section{Introduction}
\label{sec:intro}

Coherent Elastic Neutrino-Nucleus Scattering (\cvn) has proven to be a
powerful test of the Standard Model (SM) of particle physics, and a
search tool for new physics (NP).  In particular, the recent detections
of \cvn\, by COHERENT~\cite{Akimov:2017ade,Akimov:2020pdx} is able to constrain
non-standard neutrino interactions (NSI) due to heavy or light
mediators~\cite{Coloma:2017egw,Coloma:2017ncl,Liao:2017uzy,Dent:2017mpr,Kosmas:2017tsq,Billard:2018jnl,Lindner:2016wff,Abdullah:2018ykz, Farzan:2018gtr,Brdar:2018qqj},
generalized scalar and vector neutrino
interactions~\cite{AristizabalSierra:2018eqm}, and hidden sector
models~\cite{Datta:2018xty}. It also sets independent constraints on
the effective neutron size distribution of
CsI~\cite{Ciuffoli:2018qem,AristizabalSierra:2019zmy,Papoulias:2019lfi},
and on sterile neutrinos~\cite{Kosmas:2017zbh,Blanco:2019vyp}.

To this point, constraints on NP with the COHERENT data have been
obtained mostly using the measured distribution of nuclear recoil
energies. Due to the nature of the stopped-pion source utilized by
COHERENT and the detectors that are deployed, the time distribution of
events also provides a powerful probe of NP
models~\cite{Dutta:2019eml,Giunti:2019xpr}. This has proven to be
important not only in searches for NP in the neutrino sector, but also
applicable to searches for NP in the form of low-mass dark
matter~\cite{Dutta:2019nbn}.

Since the power of \cvn\, as a NP probe is just now beginning to be
realized, it is important to identify new ways to exploit \cvn\, in
future experiments. In this paper, we take a step in this direction
and investigate the prospects for supplementing the nuclear recoil
energy with the direction of the nuclear recoil. Assuming SM physics,
we calculate the expected angular distribution of nuclear recoil
events for terrestrial sources that are now being used for the
detection of \cvn. We extend to investigate the angular dependence of
\cvn\, in NP scenarios, in particular focusing on models with MeV-scale
vector or scalar mediators. 

While directional detectors are not currently being deployed for
detecting \cvn\, from terrestrial sources, research and development for
similar detectors is being actively pursued for the purpose of dark
matter detection~\cite{Mayet:2016zxu,Battat:2016pap}. Since our
analysis is primarily focused on the theoretical aspects of the energy
and directional dependence of the induced nuclear recoils, we focus on
simplified detectors models. For neutrino sources, we consider both a
stopped-pion source and a reactor source. The results that we present
are meant to guide both the theoretical and experimental efforts on
this topic.

The remainder of this paper is organized as follows. In
Section~\ref{sec:recoil-momentum-spectrum}, we review the theoretical
aspect of \cvn, laying out the formalism for the calculation of the
angular distribution of recoil events. In Section~\ref{sec:modeling},
we discuss the properties of the sources that we consider, and the
simple models for the detectors. In Section~\ref{sec:kin}, we review
some aspects of the kinematics that are important for our analysis. In
Section~\ref{sec:signatures} we make predictions for SM signatures,
and in Section~\ref{sec:SM-versus-BSM-signals} we make predictions for
NP vector and scalar mediator models.

%%%%%%%%%%%%%%%%%%%%%%%%%%%%%%%%%%%
%%%%%%%%%%%%%%%%%%%%%%%%%%%%%%%%%%%
%%%%%%%%%%%%%%%%%%%%%%%%%%%%%%%%%%%
%%%%%%%%%%%%%%%%%%%%%%%%%%%%%%%%%%%

\section{The recoil energy and directional recoil spectrum}
\label{sec:recoil-momentum-spectrum}

\cvn\, is a two-to-two process and therefore the scattering cross
section depends only on a single degree of freedom. This is often
chosen as the recoil energy, a convenient choice for most experimental
designs. The differential event rate as a function of the recoil
energy $dR/dE_r$, or the recoil spectrum (RS) for short, can
be expressed as follows:
\begin{equation}
  \label{eq:recoil-energy-spectrum}
  \frac{dR}{dE_r}=\mathcal{N}\int_{E_\nu^\text{min}}^{E_\nu^\text{max}}
  \,\frac{d\sigma}{dE_r}\,F^2(E_r)\,\frac{d\Phi}{dE_\nu}dE_\nu\ ,
\end{equation}
where $\mathcal{N}$ is the number of scattering targets,
$d\sigma/dE_r$ is the differential cross section as a function of the
recoil energy, $E_\nu$ is the incident neutrino energy, $d\Phi/dE_\nu$
is the neutrino flux, and $F(E_r)$ is the nuclear form factor. We use
the Helm form factor~\cite{Lewin:1995rx} given by\footnote{Any other
  choice as well as accounting for different proton and neutron
  distributions through indepedent proton and neutron form factors
  will have only a percent level effect, in particular for light
  nuclei such as those we consider here
  \cite{AristizabalSierra:2019zmy}.}
\begin{equation}
  \label{eq:helm-FF}
  F(E_r) = F_H(q)= 3 \frac{j_1(q)}{q r_n} e^{-(q s)^2/2}, 
\end{equation}
which assumes that the nucleonic distribution is determined by a
convolution of a uniform density of radius $r_n$ and a Gaussian
profile parametrized by the folding width $s$, which ``measures'' the
surface thickness. In (\ref{eq:helm-FF}) the momentum transfer is
given by $q = \sqrt{2 m_N E_r}$, $j_1(q)$ is the spherical Bessel
function of the first kind, $s = 0.9$ fm, and
$r_n = \sqrt{5/3 (R_\text{min}^2-3s^2)}$. For the targets that we consider
below, we have $R_\text{min} = 1.6755$ fm for He and $2.8976$ fm for F,
which correspond to the rms radii of their proton distributions
\cite{Angeli:2013epw}.

The SM differential cross section proceeds through a neutral current
process and is given by \cite{Freedman:1973yd,Freedman:1977xn}
\begin{equation}
  \label{eq:x-sec}
  \frac{d\sigma}{dE_r}=\frac{G_F^2m_N}{2\pi}g_V^2
  \left(2-\frac{m_NE_r}{E_\nu^2}\right)\ ,
\end{equation}
where $G_F$ is the Fermi constant,
$g_V=N(g_V^u+2g_V^u)+Z(2g_V^u+g_V^d)$, $N=A-Z$ with $A$ the nucleus
mass number, $Z$ is the atomic number, $m_N$ is the nuclear mass of
the detector material, $g_V^u=1/2-4/3\sin^2\theta_W$ and
$g_V^d=-1/2+2/3\sin^2\theta_W$. For the Fermi constant and the weak
mixing angles we use their PDG values:
$G_F=1.166\times 10^{-5}\,\text{GeV}^{-2}$, $\sin^2\theta_W=0.231$.
The latter obtained using the $\overline{\text{MS}}$ renormalization
scheme at the $m_Z$ scale~\cite{Olive:2016xmw}.

We now proceed to generalize the formalism to detectors with
directional sensitivity. Theoretically, the $E_r$ dependence in 
Eq. (\ref{eq:recoil-energy-spectrum}) can be traded with the direction
of recoil $\cos \theta_r$ converting the RS to an Angular Spectrum
(AS). In practice, however, a detector may provide a measurement of
both $E_r$ and $\cos \theta_r$ at once, so it would be more convenient
to express the scattering rate as a function of both variables:
  \begin{equation}
  \label{eq:momentum-spectrum}
  \frac{d^2R}{dE_rd\Omega_r}\ ,
\end{equation}
where $\Omega_r$ refers to the solid angle along the direction of the
recoiling nucleus with respect to the incoming neutrino direction. We
refer to this observable as the Directional Recoil Spectrum (DRS),
although the term ``Momentum Spectrum'' has been previously used in
the literature \cite{Gondolo:2002np}. To derive an expression for the
DRS we closely follow Ref. \cite{OHare:2015utx} where the incoming
neutrino energy $E_\nu$ is traded for the angle of the recoiling
nucleus. Note that if the neutrino source is mono-energetic this
procedure is superfluous; the two arguments of the resulting DRS would
be tied by a Dirac $\delta$-function.

The procedure requires some adaptation for neutrino production at the
SNS or nuclear reactors. The direction of the source has no seasonal
dependence as in \cite{OHare:2015utx} where the neutrinos produced in
the Sun whose location with respect to the Earth changes with time. We
are interested in terrestrial neutrino sources that are at rest with
respect to the detector and so the neutrino flux can be written as
\begin{equation}
  \label{eq:neutrino-flux-directionality}
  \frac{d^2\Phi}{dE_\nu d\Omega_\nu}=\frac{d\Phi}{dE_\nu}
  \delta(\hat q_\nu - \hat q_\text{det})\ ,
\end{equation}
where the unit vector $\hat q_\text{det}$ points from the source to
the detector while $\hat q_\nu$ defines the direction of the incoming
neutrino. Strictly speaking this expression should be thought of as
per event since both the source and detector are extended objects.

\begin{figure*}
  \centering
  \includegraphics[scale=0.77]{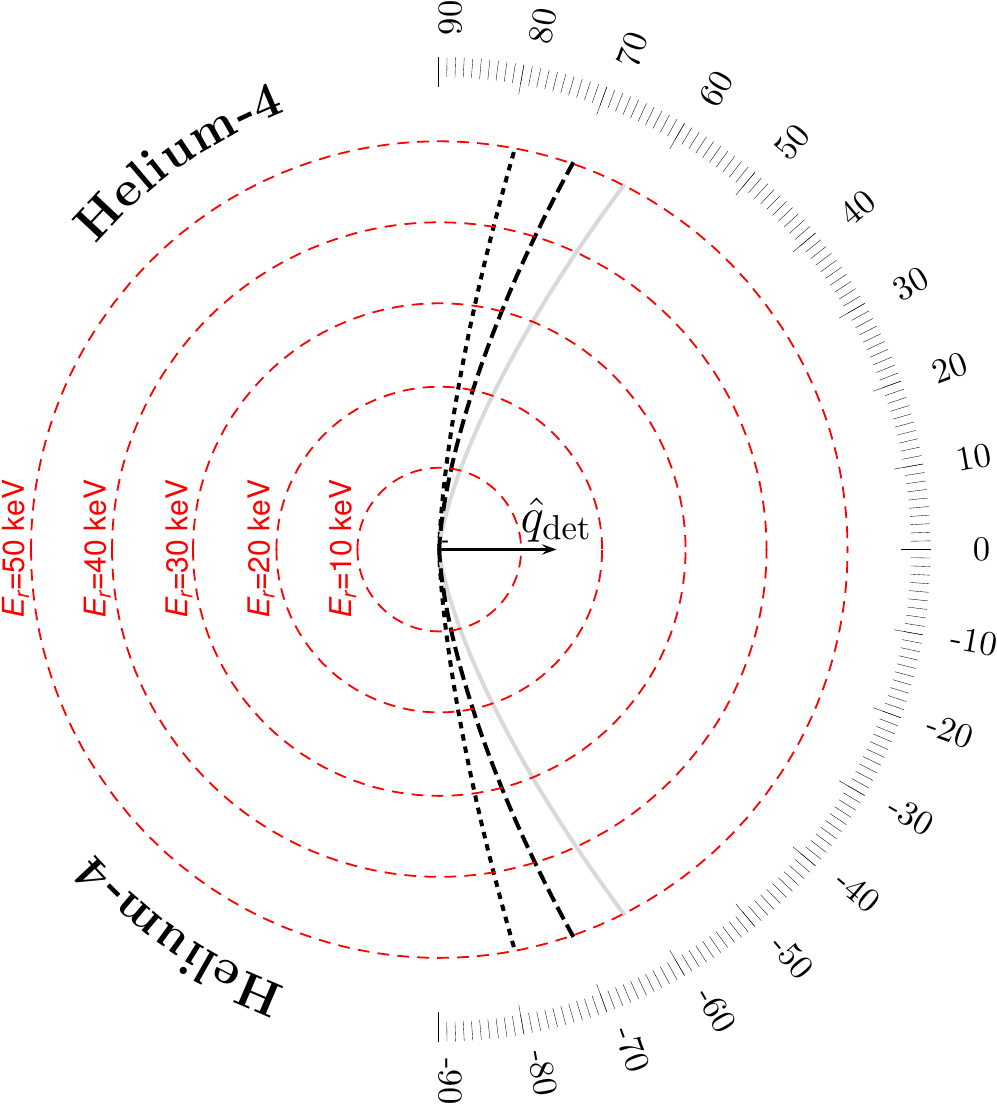}
  \hfill
  \includegraphics[scale=0.77]{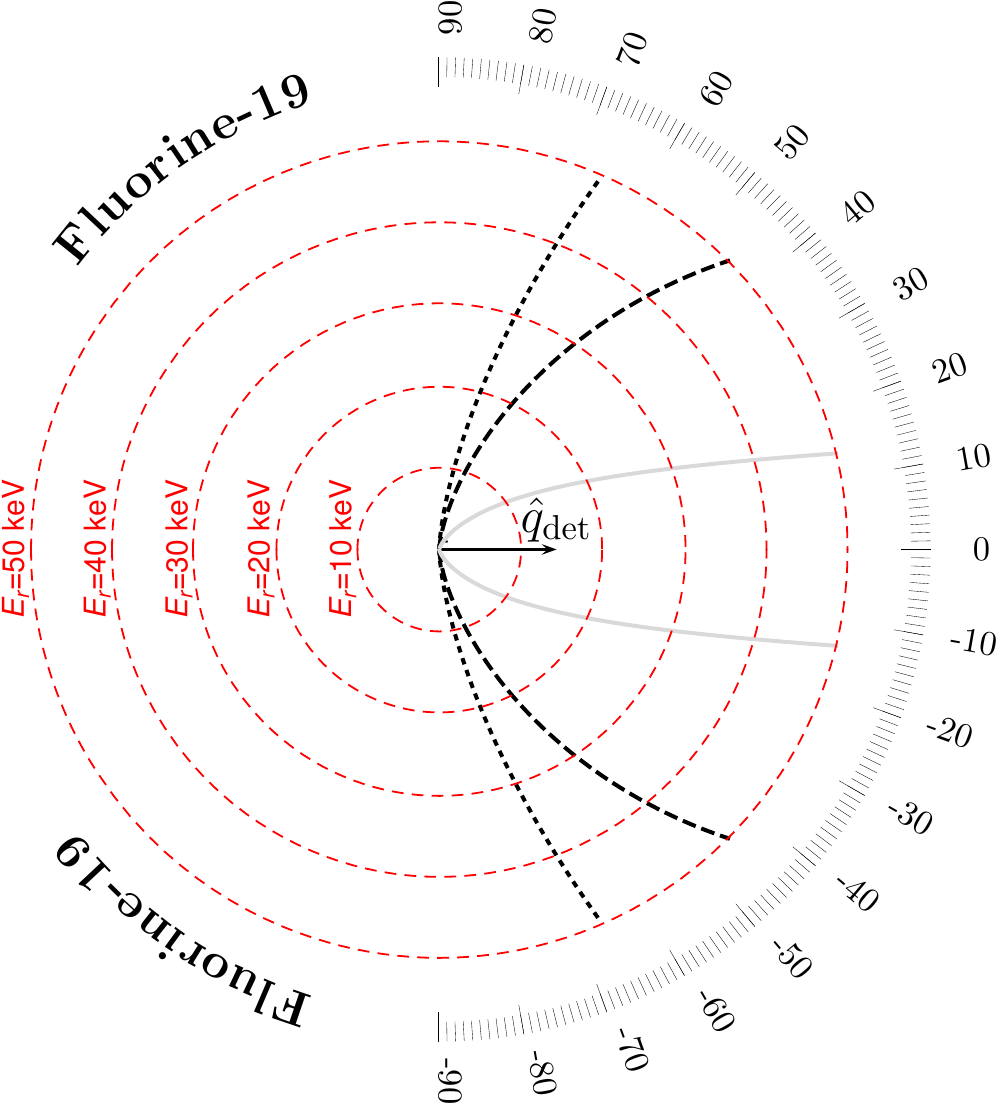}
  \caption{{\bf Left}: Allowed nuclear recoil angular region for a
    particular incoming neutrino direction (determined by the unit
    vector $\hat q_\text{det}$ which points from the neutrino source
    to the detector) for SNS and a helium detector. We include
    $\theta_r\to -\theta_r$ % (or more accurately $\phi \to \phi+\pi$)
    for illustration.  The dotted black curve is determined by the
    kinematic constraint $E_\nu=\varepsilon=m_\mu/2$ enforced by
    energy conservation and the neutrino production mechanism. The
    grey solid curve is the single event threshold
    $d^2R/dE_rd\Omega_r\geq 1$, assuming an exposure of one
    ton-yr. The dashed black curve is the angular position of the
    $\nu_\mu$ events due to the mono-energetic neutrinos (see
    Eq. (\ref{eq:numu-induced-events})). The red dashed lines are
    contours of equal recoil energy. Given an incoming neutrino
    direction the measurable angular distribution lies to the right of
    the dotted black curves. {\bf Right}: Same as the left graph but
    for a fluorine detector.}
  \label{fig:sm-momentum-spectrum-1}
\end{figure*}
In deriving the cross section in Eq. (\ref{eq:x-sec}) a 4-dimensional
$\delta$-function is evaluated completely. Here we take a step back
and leave the energy component of that $\delta$-function that relates
the incoming neutrino energy $E_\nu$ with $E_r$. The result is
\begin{equation}
  \label{eq:double-diff-xsec-interm}
  d^2\sigma=\frac{1}{64\pi^2}\frac{1}{E_\nu\,m_N}
  \frac{p_N' dE_N'd\Omega_r}{E_\nu'}\delta(E_\nu'+E_N'-E_\nu-m_N)
  |\mathcal{M}|^2. 
\end{equation}
Here we have used for the relative velocity $v_\text{rel}=1$ and the
primed (unprimed) kinematic variables refer to outgoing (ingoing)
states. Three-momentum conservation combined with energy conservation
$E_r=E_N'-m_N$ allows us to write the argument of the
$\delta$-function as a function of $\cos\theta_r$, where the nucleus
recoil angle $\theta_r$ is measured with respect to the incoming
neutrino direction, i.e.
$\cos\theta_r=\hat q_\text{det} \cdot\hat q_r$:
\begin{equation}
  \label{eq:Delta-argument}
  f(\cos\theta_r)\equiv E_r+\sqrt{E_\nu^2+p_N^{\prime 2}-
    2 E_\nu p_N^\prime\cos\theta_r}-E_\nu\ .
\end{equation}
Using the $\delta$-function identity
\begin{equation}
  \label{eq:Delta-costhetar}
  \delta[f(\cos\theta_r)]=
  \frac{\delta(\cos\theta_r-\cos\bar\theta_r)}
  {\left|df(\cos\theta_r)/d\cos\theta_r\right|}\ ,
\end{equation}
with $\cos\bar\theta_r=(m_N+E_\nu)/E_\nu\sqrt{E_r/(2m_N+E_r)}$, the
root of the equation $f(\cos\theta_r)=0$, we arrive at a rather
simplified expression for the double differential cross section
\cite{OHare:2015utx}
\begin{equation}
  \label{eq:double-diff-x-sec}
  \frac{d^2\sigma}{dE_rd\Omega_r}=\frac{1}{2\pi}\frac{d\sigma}{dE_r}
  \delta(\cos\theta_r-\cos\bar\theta_r)\ .
\end{equation}

The DRS in (\ref{eq:momentum-spectrum}) can now be
written as
 \begin{equation}
  \label{eq:momentum-spectrum-explicit-1}
  \frac{d^2R}{dE_rd\Omega_r}=\mathcal{N}\int\frac{d^2\sigma}{dE_rd\Omega_r}\,
  F^2(E_r)\frac{d^2\Phi}{dE_\nu d\Omega_\nu}\,dE_\nu d\Omega_\nu\
\end{equation}
which with the aid of Eqs. (\ref{eq:neutrino-flux-directionality}) and
(\ref{eq:double-diff-x-sec}) becomes
 \begin{equation}
  \label{eq:momentum-spectrum-explicit-2}
  \frac{d^2R}{dE_rd\Omega_r}=\frac{\mathcal{N}}{2\pi}
  \int\frac{d\sigma}{dE_r}\,
  F^2(E_r)\frac{d\Phi}{dE_\nu}\,
  \delta(\hat q_r\cdot\hat q_\text{det}-\cos\bar\theta_r)dE_\nu\ .
\end{equation}
To perform the integration we rewrite the argument of the $\delta$-function as
\begin{equation}
  \label{eq:arg-delta-fun}
  \hat q_r\cdot\hat q_\text{det}-\cos\bar\theta_r=
  E_\nu^\text{min}\left(x+\frac{1}{\varepsilon}\right)\ ,
\end{equation}
with the new variables defined by
\begin{equation}
  \label{eq:new-var}
  \frac{1}{\varepsilon}=\frac{\hat q_r\cdot \hat q_\text{det}}
  {E_\nu^\text{min}}-\frac{1}{m_N}\ ,
  \quad
  x=-\frac{1}{E_\nu}\
\end{equation}
and we used $E_\nu^\text{min} = \sqrt{m_NE_r/2}$. Integration over $x$ yields the following analytical expression for
the DRS
\begin{equation}
  \label{eq:momentum-spectrum-3}
  \frac{d^2R}{dE_rd\Omega_r}=\frac{\mathcal{N}}{2\pi}
  \left .\frac{d\sigma}{dE_r}\right|_{E_\nu=\varepsilon}\,
  F^2(E_r)\frac{\varepsilon^2}{E_\nu^\text{min}}
  \left .\frac{d\Phi}{dE_\nu}\right|_{E_\nu=\varepsilon}\ .
\end{equation}
Dependence on the nucleus scattering angle is encoded in $\varepsilon$
through $\hat q_r\cdot \hat q_\text{det}=\cos\theta_r$. 
%The momentum
%spectrum is therefore an even function of $\theta_r$.
\begin{figure*}
  \centering
  \includegraphics[scale=0.37]{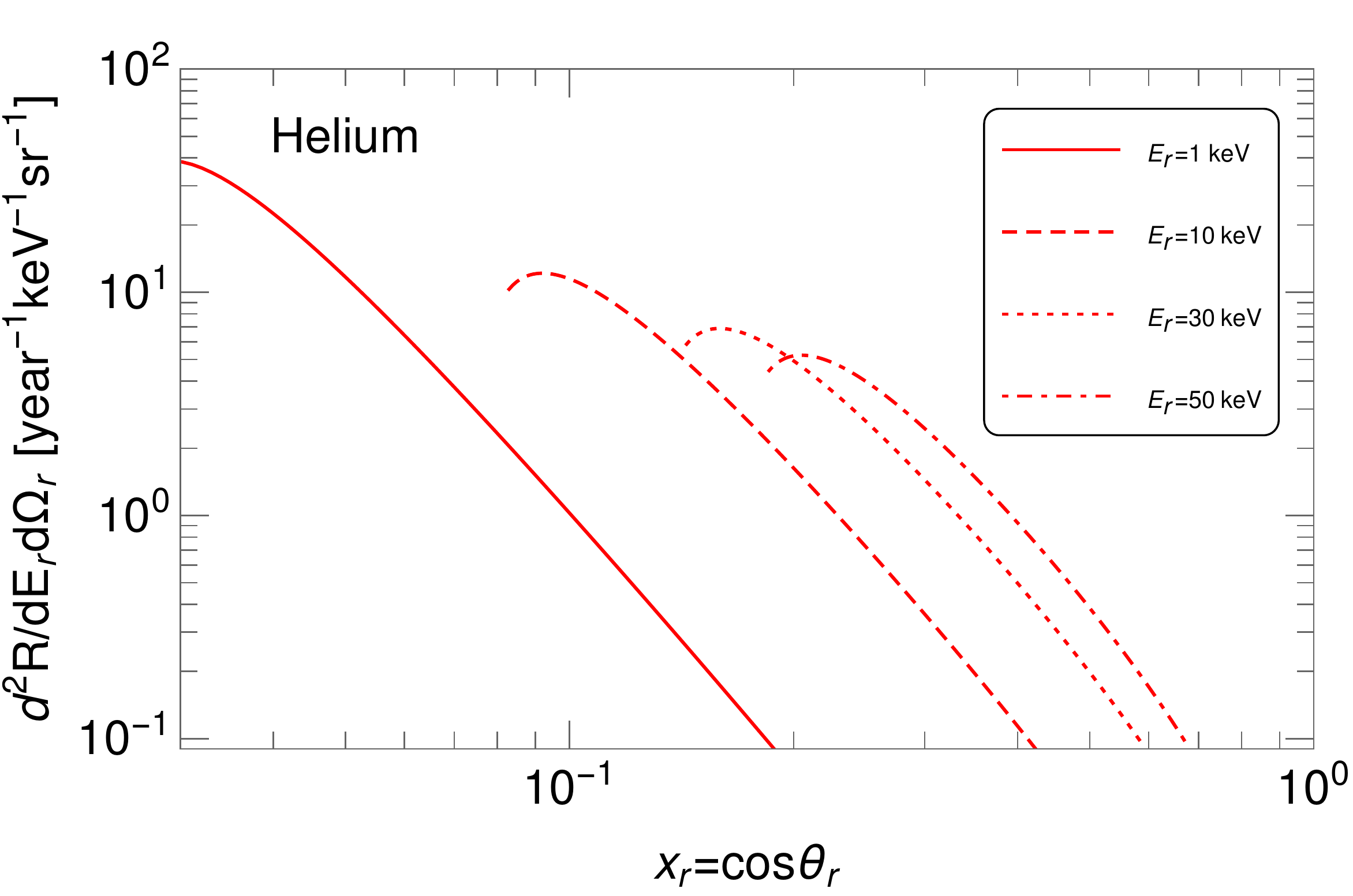}
  \hfill
  \includegraphics[scale=0.37]{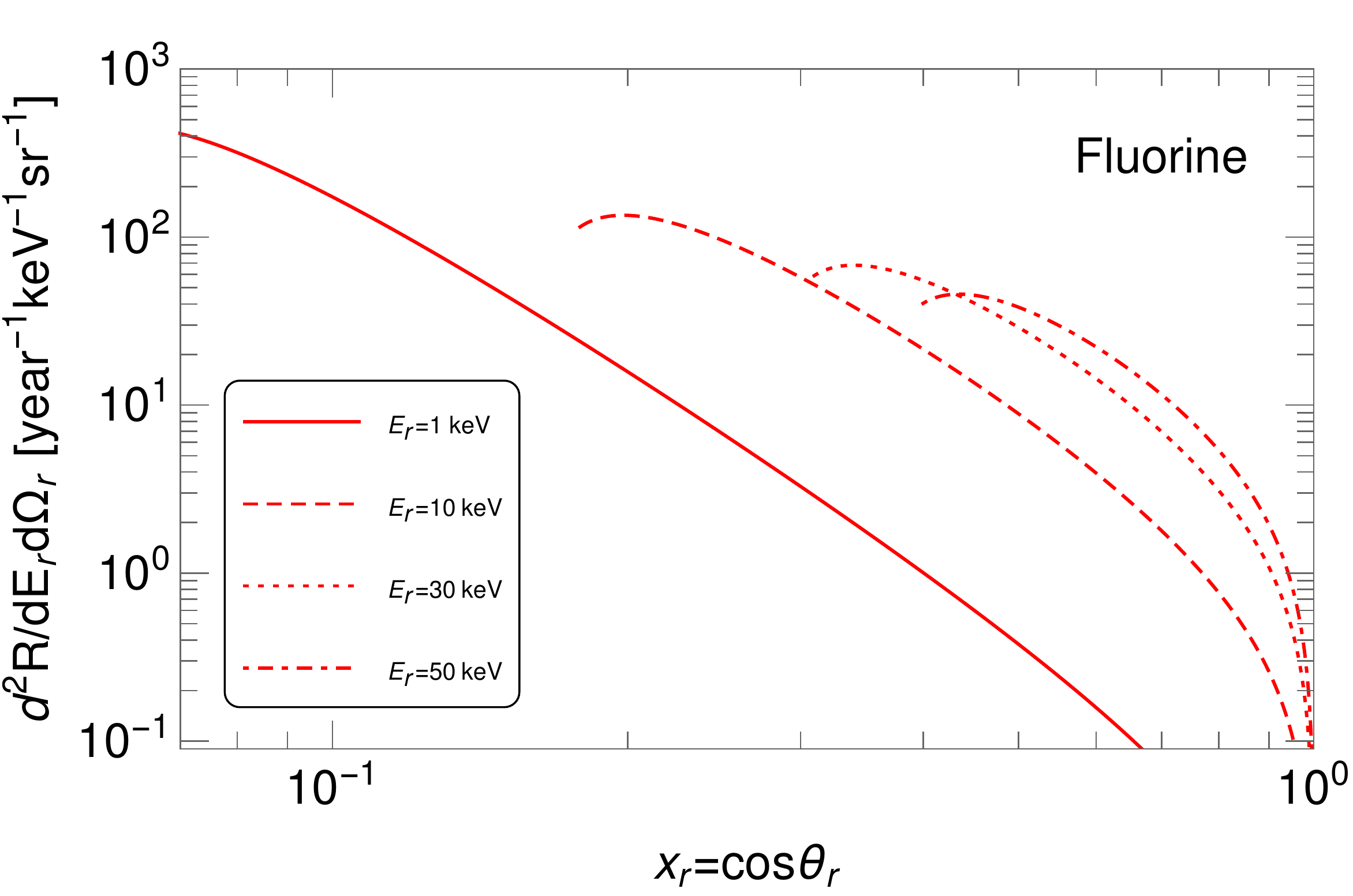}
  \includegraphics[scale=0.37]{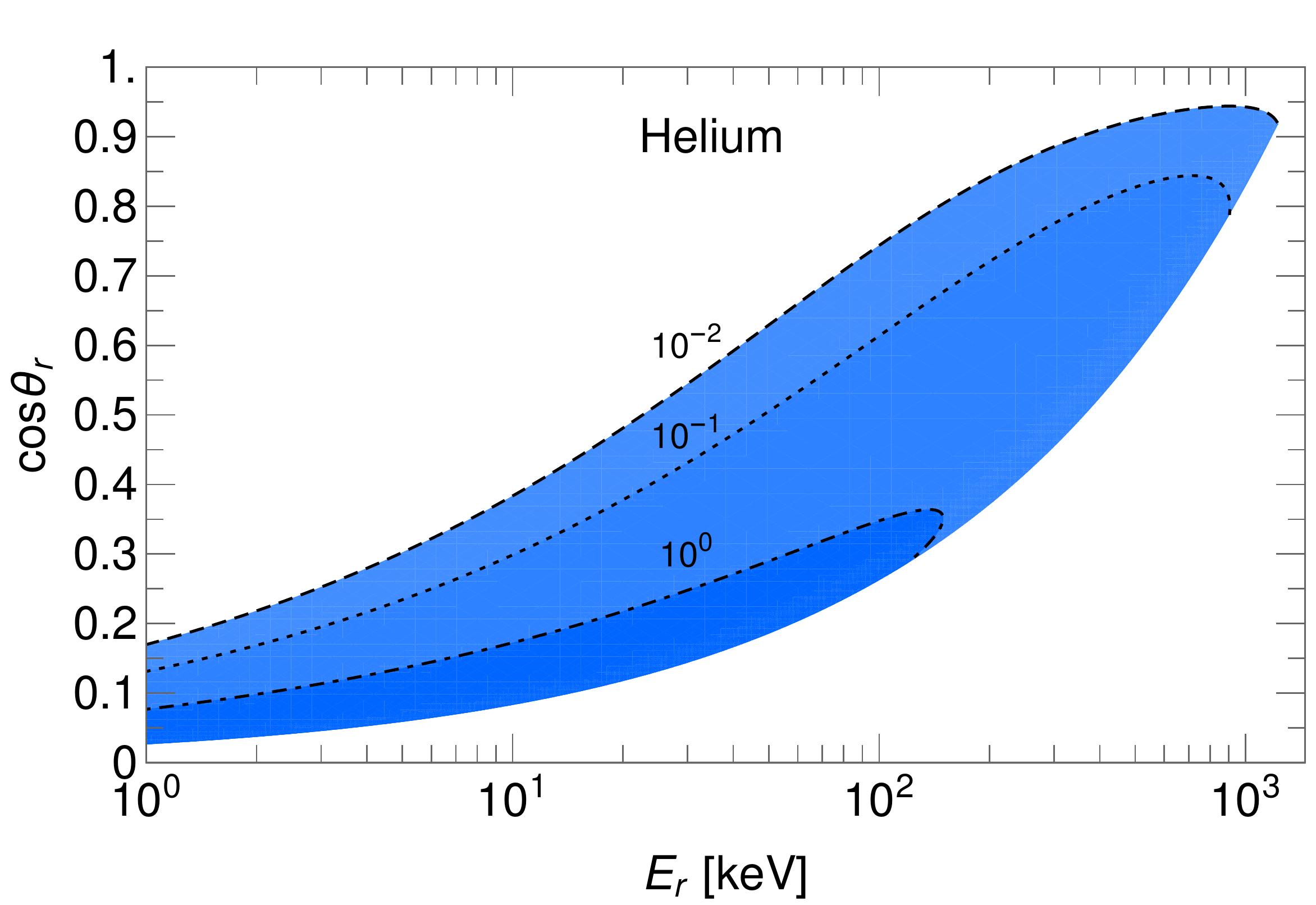}
  \hfill
  \includegraphics[scale=0.37]{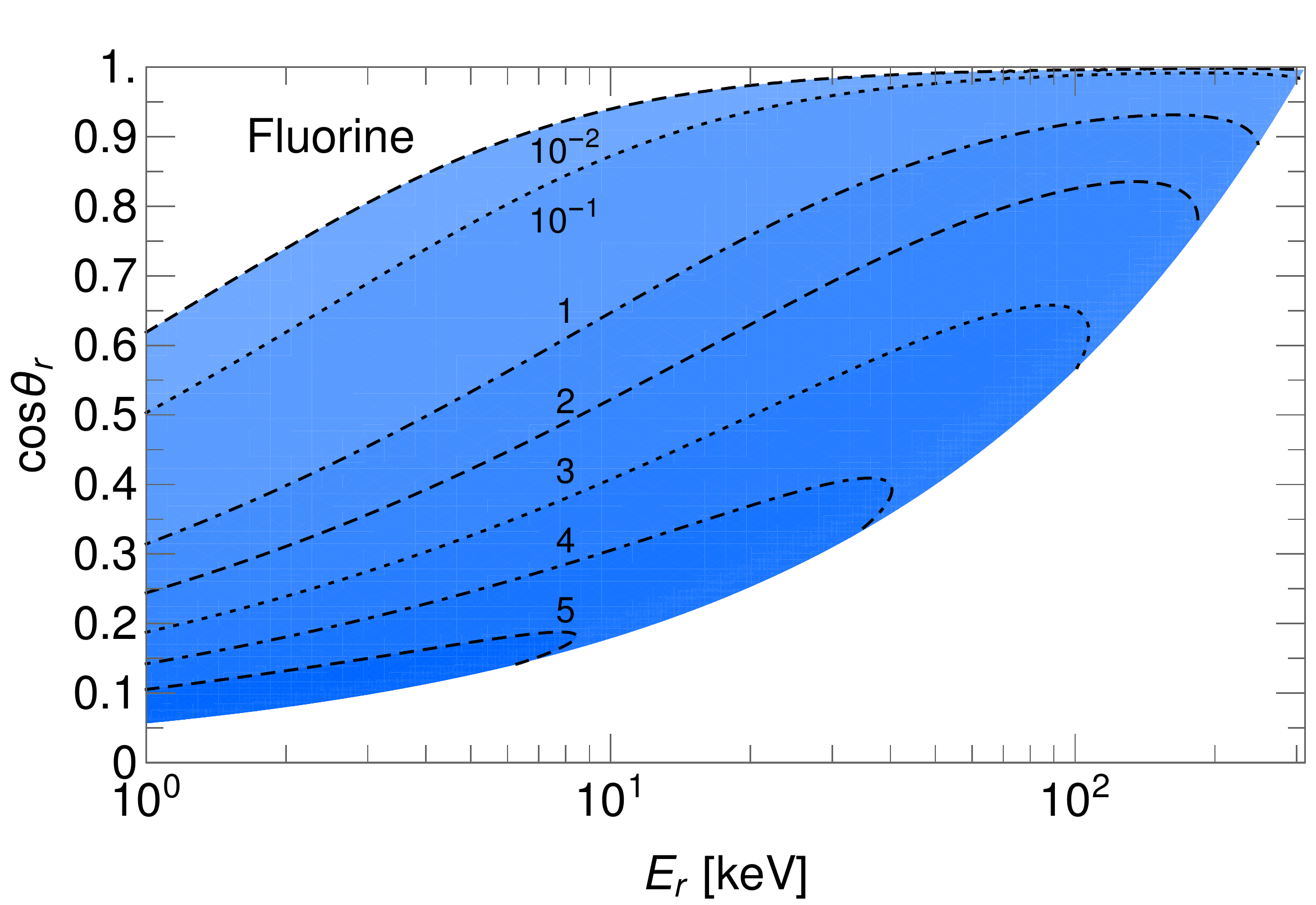}
  \caption{{\bf Top}: Nuclear recoil energy, $E_r$, slices of the DRS
    as a function of \cosr\, for He (left) and F (right) detectors.
    The DRS curves are limited on the left by the maximum neutrino
    flux energy. {\bf Bottom}: Contours of the same DRS in the
    \cosr-$E_r$ plane.}
  \label{fig:sm-momentum-spectrum}
\end{figure*}

%%%%%%%%%%%%%%%%%%%%%%%%%%%%%%%%%
%%%%%%%%%%%%%%%%%%%%%%%%%%%%%%%%%
%%%%%%%%%%%%%%%%%%%%%%%%%%%%%%%%%
%%%%%%%%%%%%%%%%%%%%%%%%%%%%%%%%%
%%%%%%%%%%%%%%%%%%%%%%%%%%%%%%%%%

\section{Source and detector modeling}
\label{sec:modeling}
As emphasized above we are interested in understanding the basic
physics of directionality in \cvn\, and will, therefore, take a
simplified approach in modeling the neutrino sources and detectors.

\subsection{Neutrino sources}
For the pion source we will assume the setup similar to that of the
COHERENT experiment at the Oak Ridge National Laboratory. The
neutrinos are produced at the Spallation Neutrino Source (SNS) by
stopped pion decays (prompt $\nu_\mu$) and consequent $\mu^+$
decays (delayed $\nu_e$ and $\bar\nu_\mu$). Thus the neutrino flux
consists of a monochromatic neutrino line at
$E_\nu=(m_\pi^2-m_\mu^2)/2 m_\pi\simeq 30\,$ MeV and two continuous
spectra. The spectral functions are given by
\begin{align}
  \label{eq:spectra}
  \mathcal{F}_{\nu_\mu}(E_\nu)&=\frac{2m_\pi}{m_\pi^2-m_\mu^2}
                                \delta\left(1-\frac{2E_\nu m_\pi}{m_\pi^2-m_\mu^2}\right)\ ,
                                \nonumber\\
  \mathcal{F}_{\nu_e}(E_\nu)&=\frac{192}{m_\mu}\left(\frac{E_\nu}{m_\mu}\right)^2
                              \left(\frac{1}{2}-\frac{E_\nu}{m_\mu}\right)\ ,
                              \nonumber\\
  \mathcal{F}_{\bar\nu_\mu}(E_\nu)&=\frac{64}{m_\mu}\left(\frac{E_\nu}{m_\mu}\right)^2
                                    \left(\frac{3}{4}-\frac{E_\nu}{m_\mu}\right)\ .
\end{align}
For a pion-at-rest source $E_\nu^\text{max}=m_\mu/2$ where
$m_\mu = 105.65$ MeV is the muon mass \cite{Olive:2016xmw}. The
neutrino flux is then obtained by normalizing these spectral functions
to $n_\text{POT}\times r/4 \pi L^2$, where $n_\text{POT}$ refers to
the number of protons at target ($1.76\times 10^{23}$ over 308.1
live-days of neutrino detection for the COHERENT CsI detector
\cite{Akimov:2017ade}), $r=0.08$ is the number of neutrinos produced
per proton-mercury collision and $L=20$ m is the detector location
from the collision point. To convert the exposure time to a whole year
we scale $n_\text{POT}$ by 365/308.1.

As for reactors, we use the Kopeikin neutrino spectral data points
\cite{Kopeikin:2012zz} normalized under the assumption of 6
anti-neutrinos and 200 MeV of energy per fission on average. Assuming
a generic $1$ GW reactor with an isotropic flux at a distance
$L/\text{cm}$ from the detector we estimate the number of neutrinos to
be
\begin{equation}
  \label{eq:reactor-normalization}
  n_\text{reactor}(L)=\frac{1.5\times 10^{19}}{(L/\text{cm})^2}
    \,\text{cm}^{-2}\,\text{s}^{-1}\ .
\end{equation}
It is important to note that both sources are modeled as
point-like. This is not a big issue for energy measurement, but for an
angular measurement using the actual size of the source leads to an
irreducible uncertainty of roughly the angular size of the source as
viewed from the detector. For example, a 1 GW reactor core of 4 m
height and 3 m diameter at a distance of 20 m has an angular size of
about 10$^\circ$. We continue with this point-like source
approximation, and discuss below the impact of this assumption.
% -------------
% Subsection
% -------------
\subsection{Detectors}
We will restrict ourselves to helium (He) and fluorine (F) detectors
which, given their natural isotopes abundances, are mainly composed of
$^4$He and $^{19}$F. We consider F because it is a standard gas used
for directional dark matter detection, and consider He because it
gives us an example of a very light nuclear target. For concreteness
we will assume a useable (fiducial) detector mass of 1 tonne located
20 m away from the source. A 1000 m$^3$ detector at normal temperature
and pressure amounts to about 164 kg of He and 1555 kg of F. In
reality, however, in a drift chamber with directional sensitivity the
target gas is at a partial pressure of about 1/75 and somewhere
between 10\% and 40\% of the mass is not useable \footnote{Private communication
with Neil Spooner and Sven Vahsen}.

We assume the detectors to have 100\% efficiency, perfect energy and
angular resolution, and do not model any backgrounds since we are
interested purely in the signal. In reality, the efficiency is
expected to deteriorate at small $E_r$ and the angular resolution can
vary from 10$^{\circ}$ to 60$^{\circ}$ and is often at the expense of
energy resolution. We also assume the detectors to be point-like or,
equivalently, to have perfect resolution of the location of the
scattering event. Unless explicitly stated otherwise, we
assume a minimum energy detection threshold of 1 keV.
\begin{figure*}
  \centering
  \includegraphics[scale=0.37]{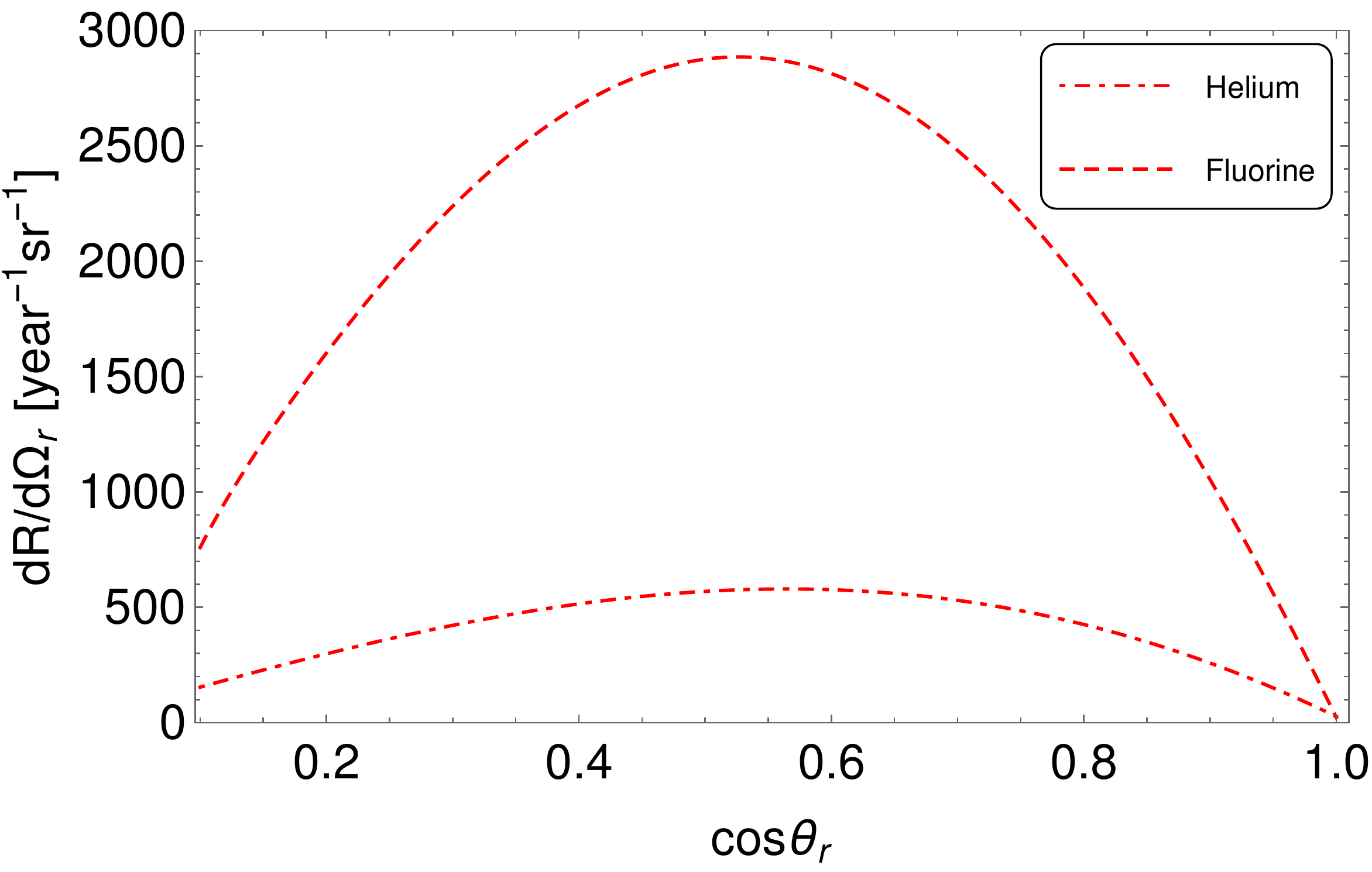}
  \hfill
  \includegraphics[scale=0.37]{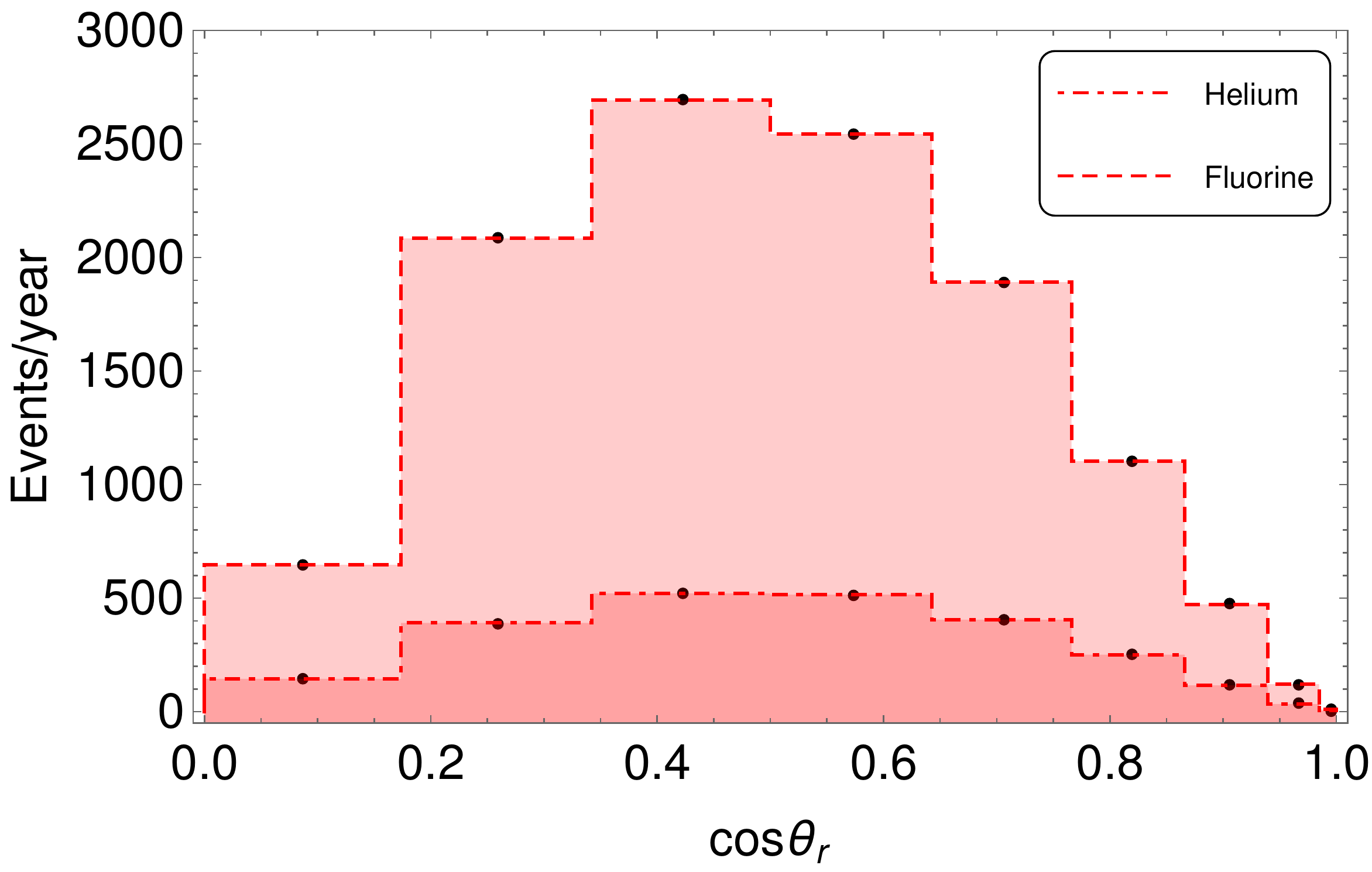}
  \caption{{\bf Left}: The angular spectra of SNS neutrinos for He and
    F detectors in the SM. The peaks occur at
    $\cos\theta_r\simeq 0.56$ for He and $\cos\theta_r\simeq 0.51$ for
    F, which translates into $\theta_r\simeq 56^\circ$ and
    $\theta_r\simeq 59^\circ$ respectively. {\bf Right}: The event
    yield per year in angular bins of size of
    $|\Delta\theta_r|=10^\circ$. The total yield is roughly 2300
    events for He and 11200 for F.}
  \label{fig:sm-recoil-angular-spectrum}
\end{figure*}
% ------------
% Section
% ------------
\section{Kinematics}
\label{sec:kin}
We now move on to discuss kinematic limits applicable to our
analysis. The recoil energy can be expressed either in terms of the
scattering angle of the neutrino, $\cos\theta$, or the nucleus,
$\cos\theta_r$. In the laboratory frame they read
\begin{align}
  \label{eq:recoil-energy}
  E_r&=\frac{E_\nu^2(1-\cos\theta)}{m_N+E_\nu(1+\cos\theta)}\ ,
       \nonumber\\
  E_r&=\frac{2m_NE_\nu^2\cos^2\theta_r}{(E_\nu+m_N)^2-E_\nu^2\cos^2\theta_r}\ .
\end{align}
From these expressions one can see that the maximum recoil energy is
obtained at forward neutrino scattering ($\theta=\pi$) and
$\theta_r=0$, while for $\theta=0$ and $\theta_r=\pi/2$ the recoil
energy vanishes. In practice, however, the maximum value for $E_r$ is
determined by the kinematics of the ingoing neutrinos, which for the
SNS is determined by $E_\nu\leq m_\mu/2$. For our reactor analysis, we
set $E_\nu\lesssim E_\nu^\text{re}=9\,$MeV. This kinematic constraint
can be translated into an upper bound on $\theta_r$ by using the
energy conservation relation $E_\nu=\varepsilon$ with
Eq.~(\ref{eq:new-var}), resulting in
\begin{alignat}{2}
  \label{eq:lower-bound-cos}
  \text{SNS}:&\quad&\cos\theta_r>\frac{1}{m_\mu}\sqrt{\frac{m_NE_r}{2}}
  \left(2 + \frac{m_\mu}{m_N}\right)\ ,
  \\\nonumber
  \text{Reactor}:&\quad&\cos\theta_r>\frac{1}{E_\nu^\text{re}}\sqrt{\frac{m_NE_r}{2}}
  \left(1 + \frac{E_\nu^\text{re}}{m_N}\right)\ .
\end{alignat}
We can see that, for a fixed recoil energy, the heavier the target
nucleus the smaller is the maximum recoil angle. For fixed nuclide
mass, larger values of recoil energy imply smaller recoil
angles. Since (\ref{eq:lower-bound-cos}) is a purely kinematic bound,
it is valid regardless of whether or not one assumes new physics
contributions.

Another constraint one could place stems from the condition
$d^2R/dE_rd\Omega_r\geq 1$, corresponding to the condition of the DRS
being measurable.  Additionally, in contrast to the kinematic limit
discussed above, this limit \textit{does} depend on the presence of
new physics. If the new contribution enhances (reduces) the DRS
\footnote{Sizable reductions are possible only for a vector
  contribution (destructive interference). Scalar interactions to a
  certain degree can destructively interfere as well, but the amount
  of reduction is proportional to either left-right neutrino mixing
  (in the case of Dirac couplings) or neutrino masses (in the case of
  Majorana couplings).} a wider (narrower) \cosr\, region can be
measured.

\begin{figure*}
  \centering
  \includegraphics[scale=0.37]{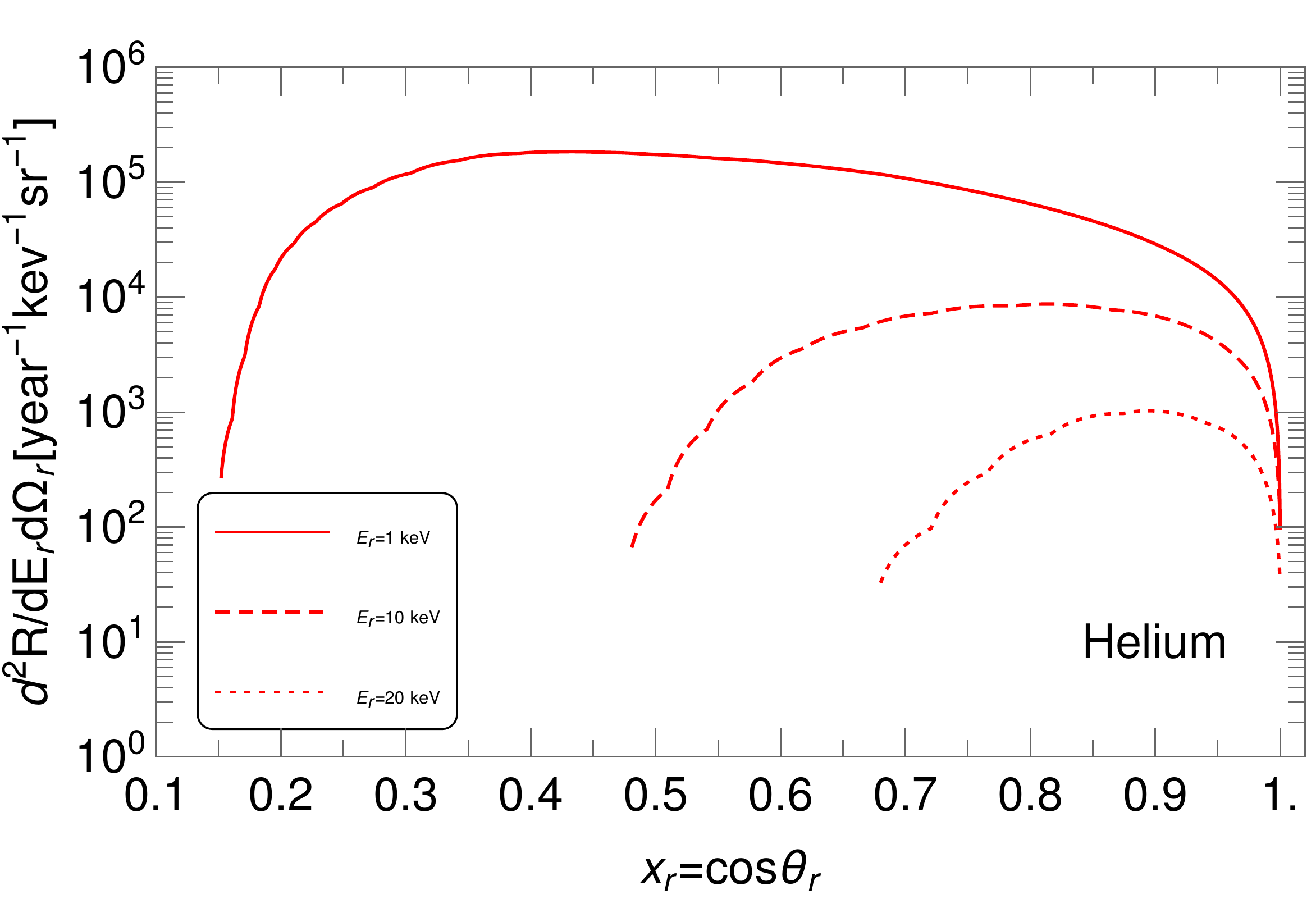}
  \hfill
  \includegraphics[scale=0.37]{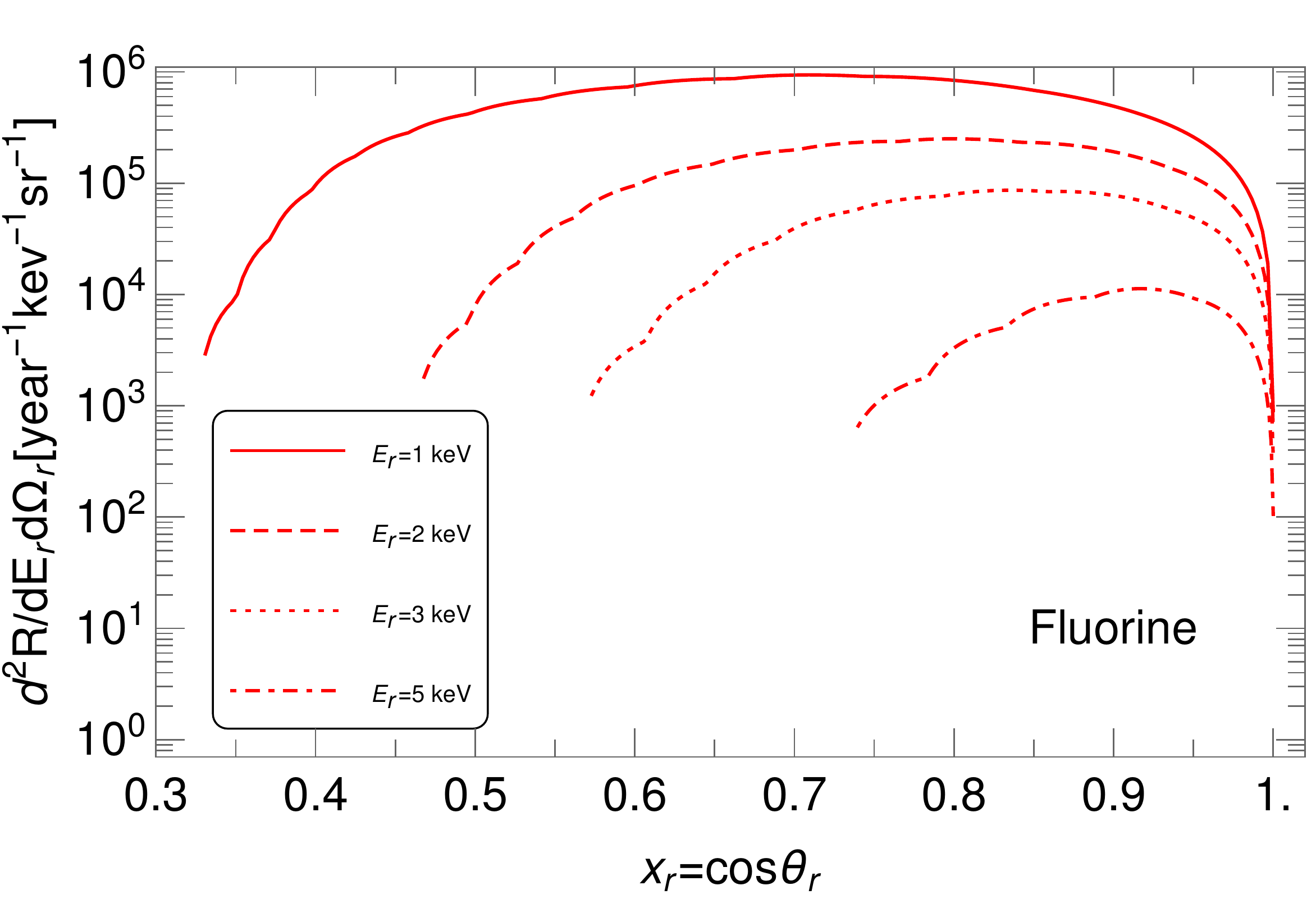}
  \caption{Nuclear recoil energy, $E_r$, slices of the DRS as a function of \cosr\, at He (left) and F
  (right) detectors from reactor neutrinos.}
  \label{fig:sm-momentum-spectrum-reactor}
\end{figure*}
The limits are illustrated in Fig. \ref{fig:sm-momentum-spectrum-1}
which shows the possible angular distributions for one-tonne helium
(left graph) and fluorine (right graph) directional detectors with SNS
neutrinos. Note that we include $\theta_r\to -\theta_r$ % (or more
% accurately $\phi \to \phi+\pi$)
for illustration. The measurable
angular region is that within the dotted and solid curves and can be
extended further towards zero degrees by increasing the exposure. One
can see that He detectors have access to larger angles than F
detectors due to the lower mass of the target. The dashed curves
correspond to the angular distribution of $\nu_\mu$-induced events. It
follows from the condition
$E_\nu=\varepsilon=(m_\pi^2-m_\mu^2)/2/m_\pi$ which translated into
$\cos\theta_r$ reads
\begin{equation}
  \label{eq:numu-induced-events}
  \cos\theta_r^{\nu_\mu}=\frac{2m_\pi}{m_\pi^2-m_\mu^2}
  \sqrt{\frac{m_NE_r}{2}}
  \left(
    1 + \frac{m_\pi^2-m_\mu^2}{2m_\pi m_N}
  \right)\ .
\end{equation}
%%%%%%%%%%%%%%%%%%%%%%%%%%%%
%%%%%%%%%%%%%%%%%%%%%%%%%%%%
%%%%%%%%%%%%%%%%%%%%%%%%%%%%

\section{Standard model Signatures}
\label{sec:signatures}

\subsection{SNS neutrinos}
\label{sec:SNS-nus}
With the aid of Eq. (\ref{eq:momentum-spectrum-3}) we can calculate
the DRS as a function of nuclear recoil angle for different recoil
energy values. Fig.~\ref{fig:sm-momentum-spectrum} shows slices of
fixed $E_r$ of the DRS and contours in the $E_r$-\cosr\, plane for
helium and fluorine. Note that we omit the prompt neutrino
contributions since they would manifest as a $\delta$-function.

Notice that F leads to markedly higher event rates and allows access
to a much larger range of energies and angles due to its mass. One
small trade off is that He can lead to larger scattering angles for
the same recoil energy. This can be seen by comparing the endpoints of
the red curves of the same energy.

Another observation is that low $E_r$ events populate regions of large
$\theta_r$ and produce substantially more events than high $E_r$.  The
contours also show that, for He, a sizable region of the DRS is within
$E_r\lesssim 100\,$keV and $\cos\theta_r\lesssim 0.3$ whereas F
results in a much wider region that spans values up to
$E_r\simeq 300\,$keV and $\cos\theta_r\simeq 0.9$. This result is
expected; smaller incoming neutrino energies induce smaller recoils
for which $\cos\theta_r\to 0$ (see Eq. ~(\ref{eq:recoil-energy})), and
around such energies the neutrino flux is more abundant. As the
incoming neutrino energy increases the recoils become more pronounced,
thus leading to larger $\cos\theta_r$ and less events due to the lower
neutrino flux.

The angular behavior can be more easily understood by examining the
angular spectrum, which can be obtained either by integrating the DRS
over $E_r$ (Eq. (\ref{eq:momentum-spectrum})) or by making a change of
variable $E_r \to \cos\theta_r$ in the recoil spectrum
(Eq. (\ref{eq:recoil-energy-spectrum})). The resulting distribution is
shown in Fig. \ref{fig:sm-recoil-angular-spectrum} both as a
continuous curve and a histogram with a bin size of
$|\Delta\theta_r|=10^\circ$. The plots show more clearly the larger
event rate in F detectors compared to He detectors, everything else
being equal. The SM cross section decreases linearly with $E_r$
(Eq. (\ref{eq:x-sec})) while the flux samples central values of
$E_r$. The combination leads to a peak around $\theta_r=56^\circ$ for
He and $\theta_r=59^\circ$ for F and a rapidly decaying distribution
at larg e \cosr, which are associated with maximum recoil
energies. Note that the curves do not extend all the way to \cosr$=0$;
they are truncated at about \cosr$=0.026$ (or 89$^\circ$) for He and
0.057 (87$^\circ$) for F due to the assumed 1 keV detector threshold
and the maximum neutrino flux energy. The one ton-year exposure yield is
about 2300 events for He and 11200 for F.
%%%%%%%%%%%%%%%%%%%%%%%%%%%%%%%%%%%%
%%%%%%%%%%%%%%%%%%%%%%%%%%%%%%%%%%%%
%%%%%%%%%%%%%%%%%%%%%%%%%%%%%%%%%%%%
\subsection{Reactor neutrinos}
\label{sec:reactor-nus}
For nuclear reactors the flux decreases almost monotonically above 1
MeV, which is the smallest accessible energy with a 1 keV detector
threshold, and becomes negligible at around 9 MeV. We cut off the flux
at around this value leading to a maximum possible $E_r$ of 43 keV for
He and 9 keV for F.

\begin{figure*}
  \centering
  \includegraphics[scale=0.37]{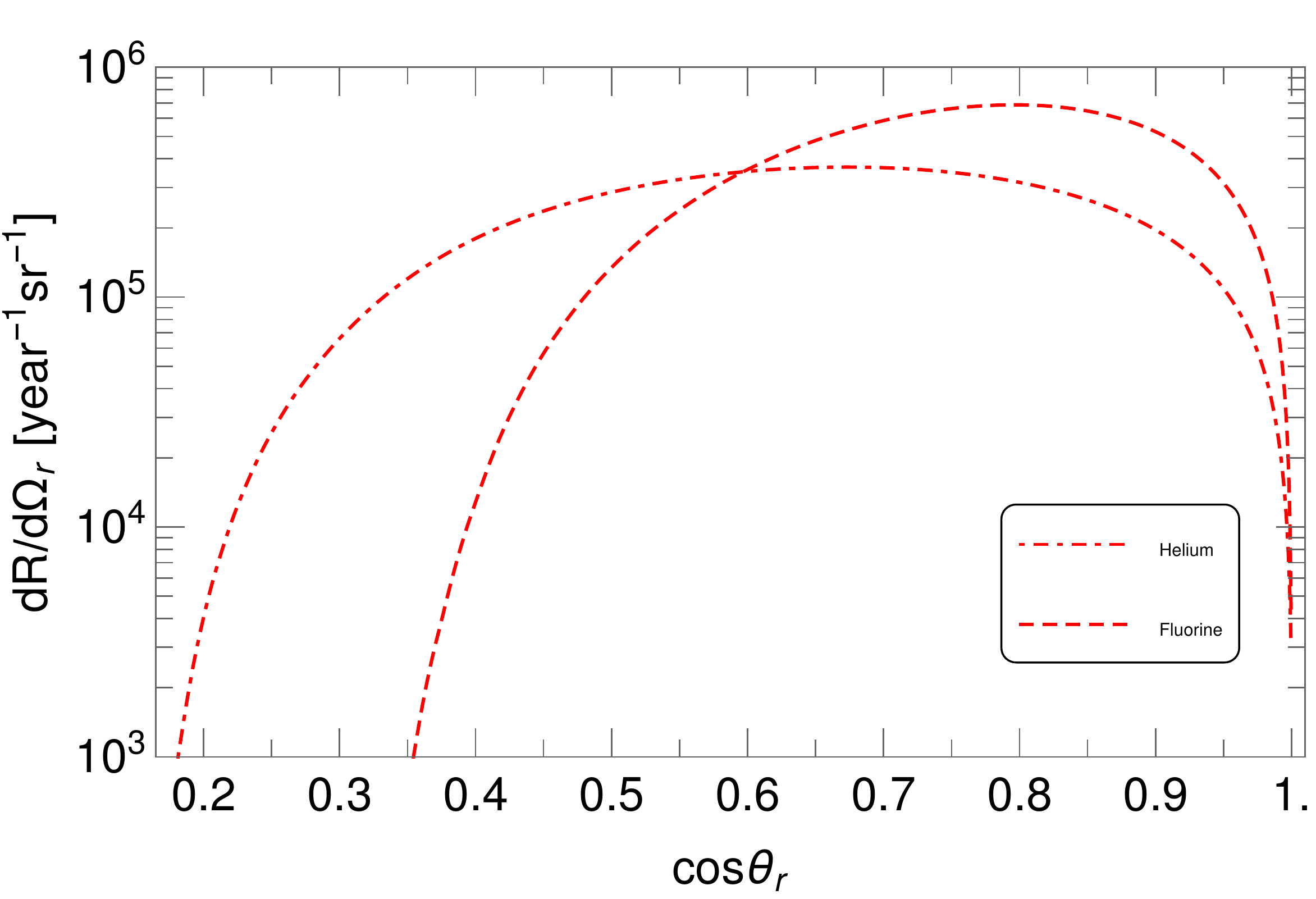}
  \hfill
  \includegraphics[scale=0.37]{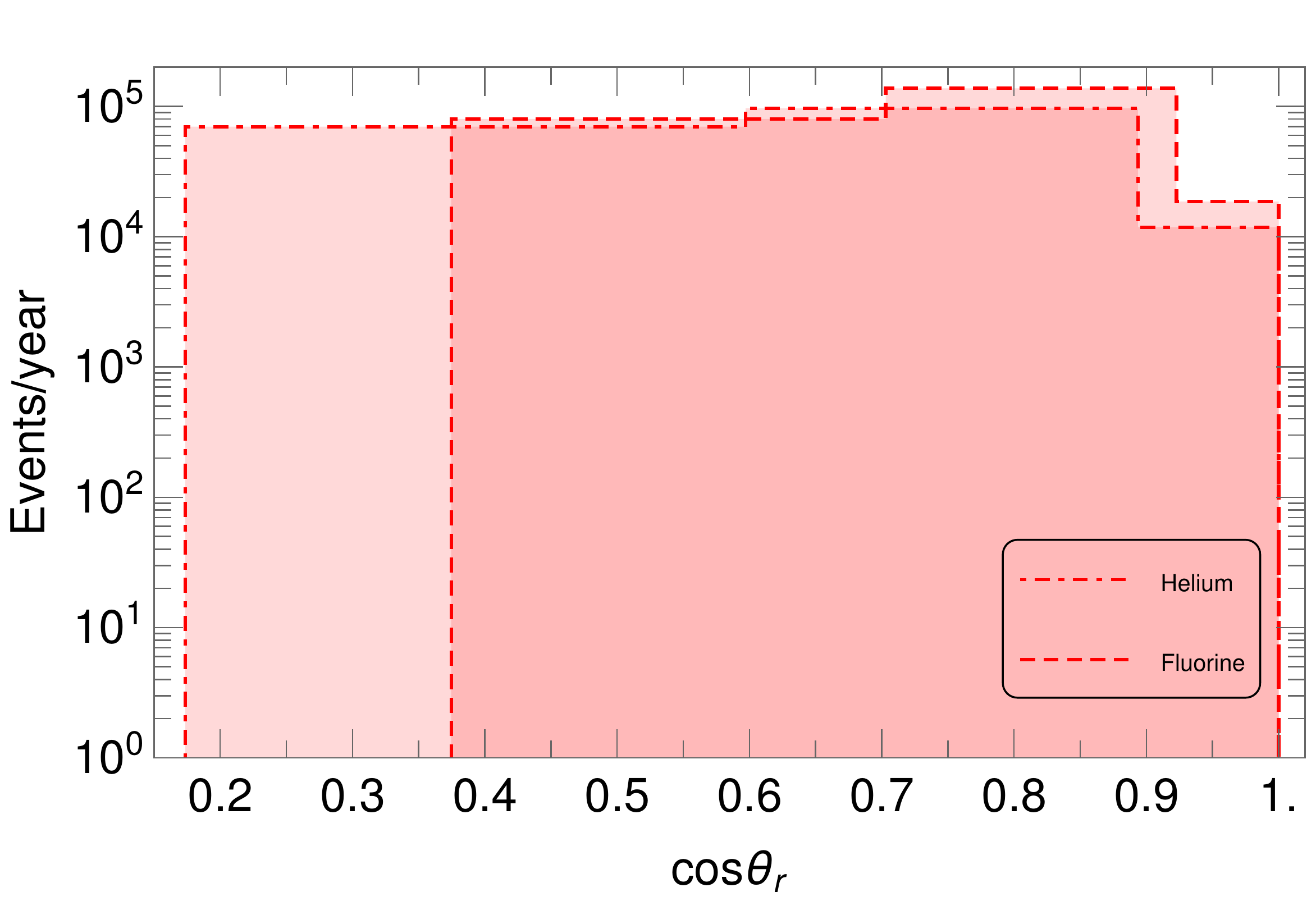}
  \caption{{\bf Left}: The angular spectra of reactor neutrinos for He
    and F detectors in the SM. {\bf Right}: The event yield per year
    in angular bins of size of $|\Delta\theta_r|=30^\circ$.}
  \label{fig:sm-recoil-angular-spectrum-reactor}
\end{figure*}

The DRS slices are shown in
Fig. \ref{fig:sm-momentum-spectrum-reactor} (note the smaller values
of $E_r$ compared to Fig. \ref{fig:sm-momentum-spectrum}). The F
detector is not able to access small \cosr\, as compared with a pion
source due to the lower maximum neutrino energy. This is seen more
clearly in Fig. ~\ref{fig:sm-recoil-angular-spectrum-reactor}
(histogram bin size is $30^\circ$) where the F distribution decays
rather quickly at around \cosr$=0.35$. On the flip side, both the He
and F curves show a remarkably larger number of total events compared
to SNS. Note, however, that a fair comparison of the two sources
requires at least accurate modeling of backgrounds and timing
information.
% -----------------
% BSM section
% -----------------
\section{New physics signatures}
\label{sec:SM-versus-BSM-signals}
\subsection{The Models}
To examine the capability of directional detectors to identify the
presence of new physics, we consider simplified models of light vector
or scalar mediators, which have been studied, for example, in Refs.
\cite{Shoemaker:2017lzs,Liao:2017uzy,Dutta:2017nht,AristizabalSierra:2017joc,Farzan:2018gtr,AristizabalSierra:2019ufd,Miranda:2020zji}.
These simplified scenarios can be accommodated in the context of gauge
invariant models, e.g., $L_\mu-L_\tau$~\cite{He:1990pn,He:1991qd},
$U(1)_{B-L}$~\cite{Heeck:2014zfa,Jeong:2015bbi,Babu:2017olk},
$U(1)_{T_{3R}}$~\cite{Dutta:2020jsy,Dutta:2019fxn},
$U(1)'$~\cite{Farzan:2015doa,Farzan:2015hkd}. Both scalar and
vector mediators can appear concurrently in the context of realistic
models.  In addition,  same type of mediator with  different masses and couplings can exist in models. Here we adopt a phenomenological approach in which only
couplings relevant for \cvn\, are
considered.
\begin{figure*}
  \centering
  \includegraphics[scale=0.37]{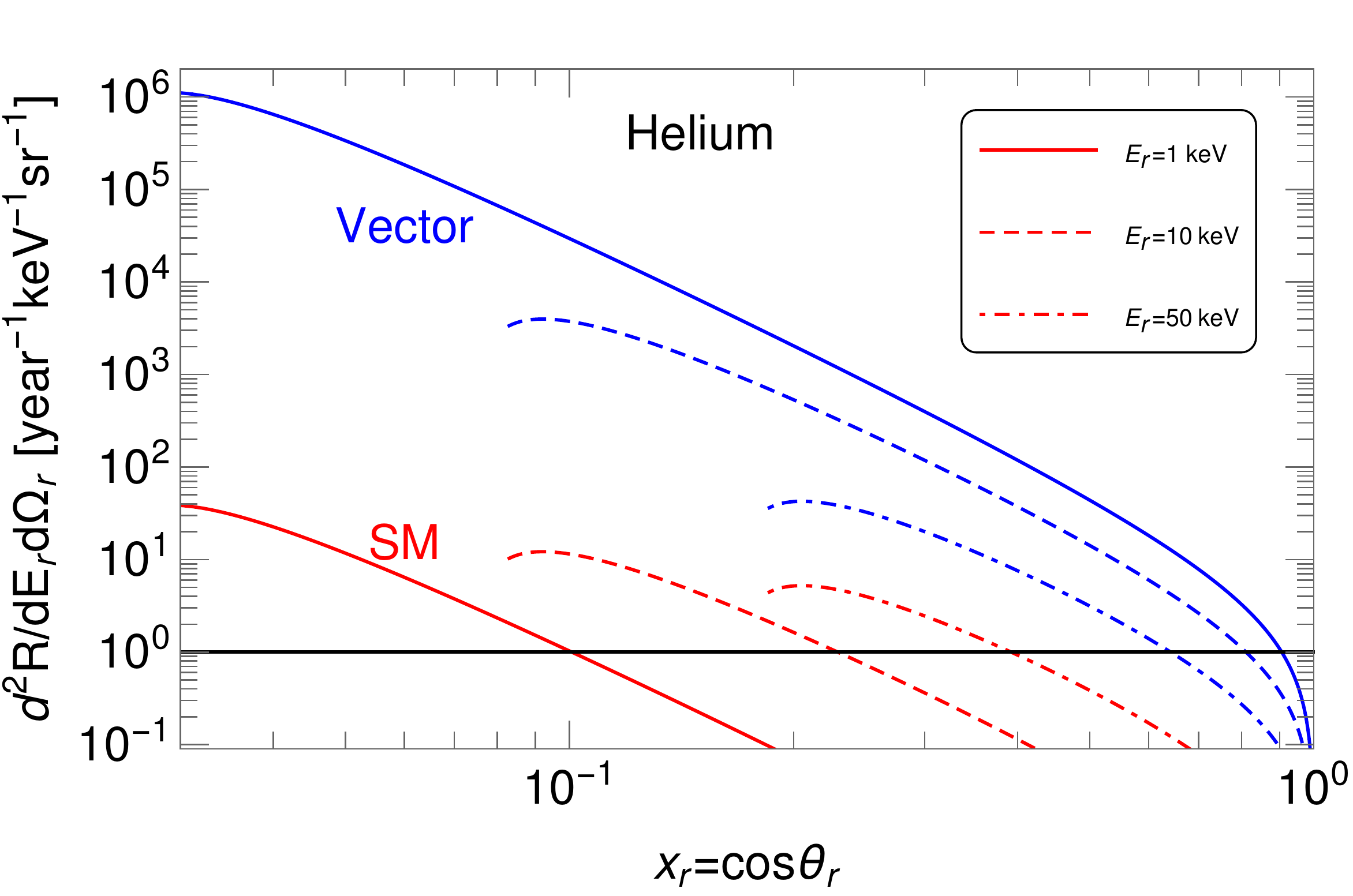}
  \hfill
  \includegraphics[scale=0.37]{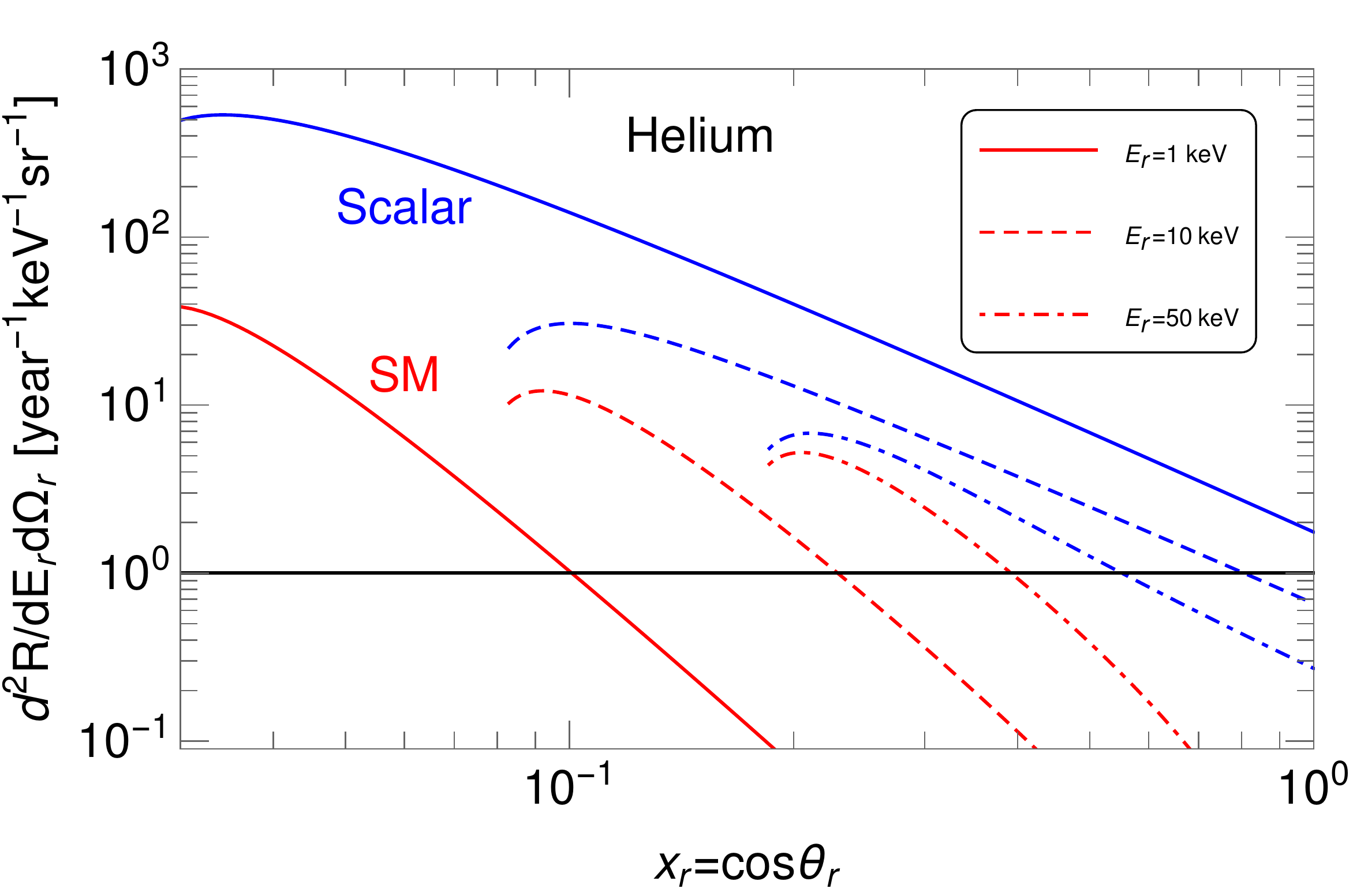}
  \includegraphics[scale=0.37]{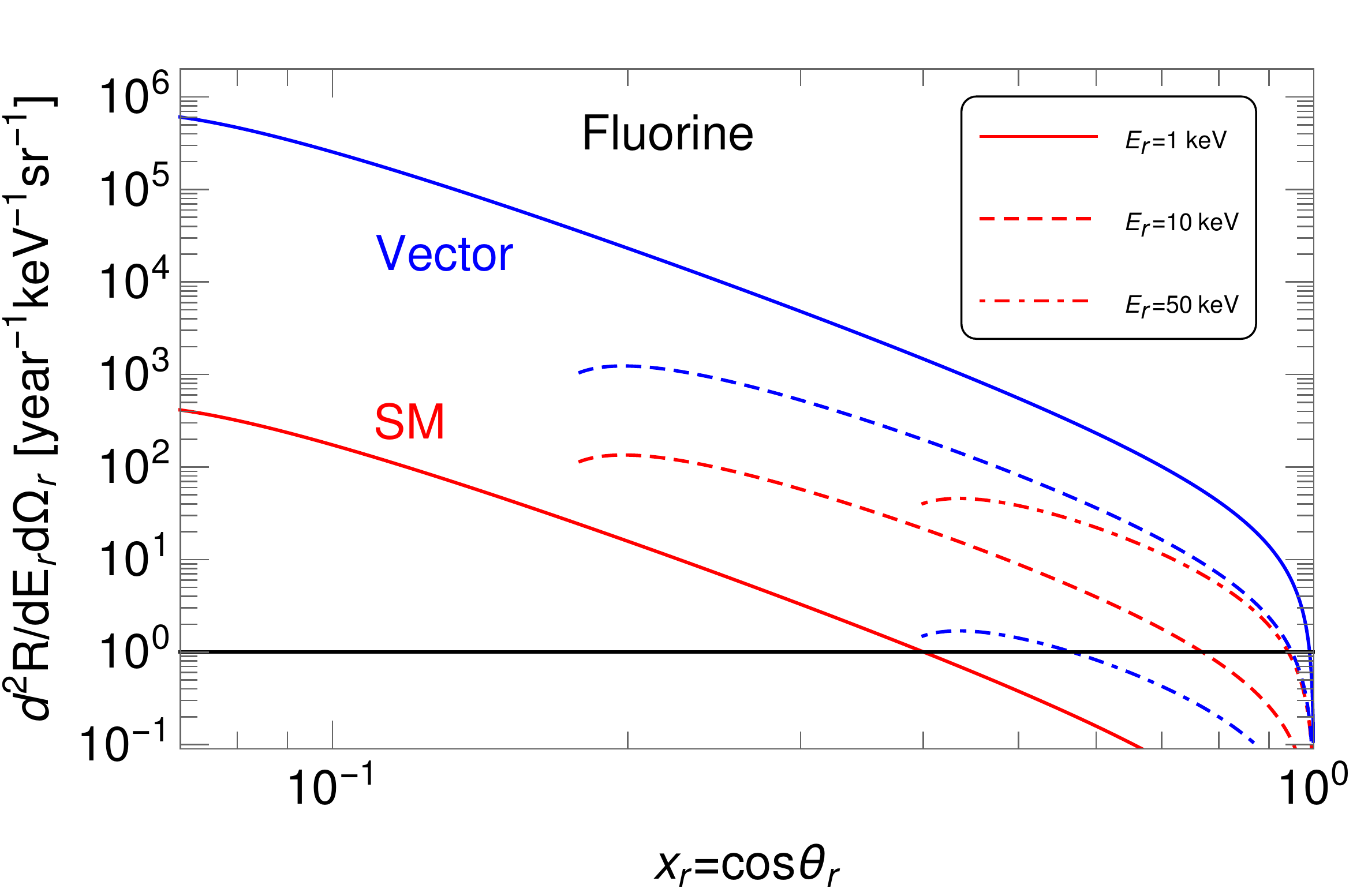}
  \hfill
  \includegraphics[scale=0.37]{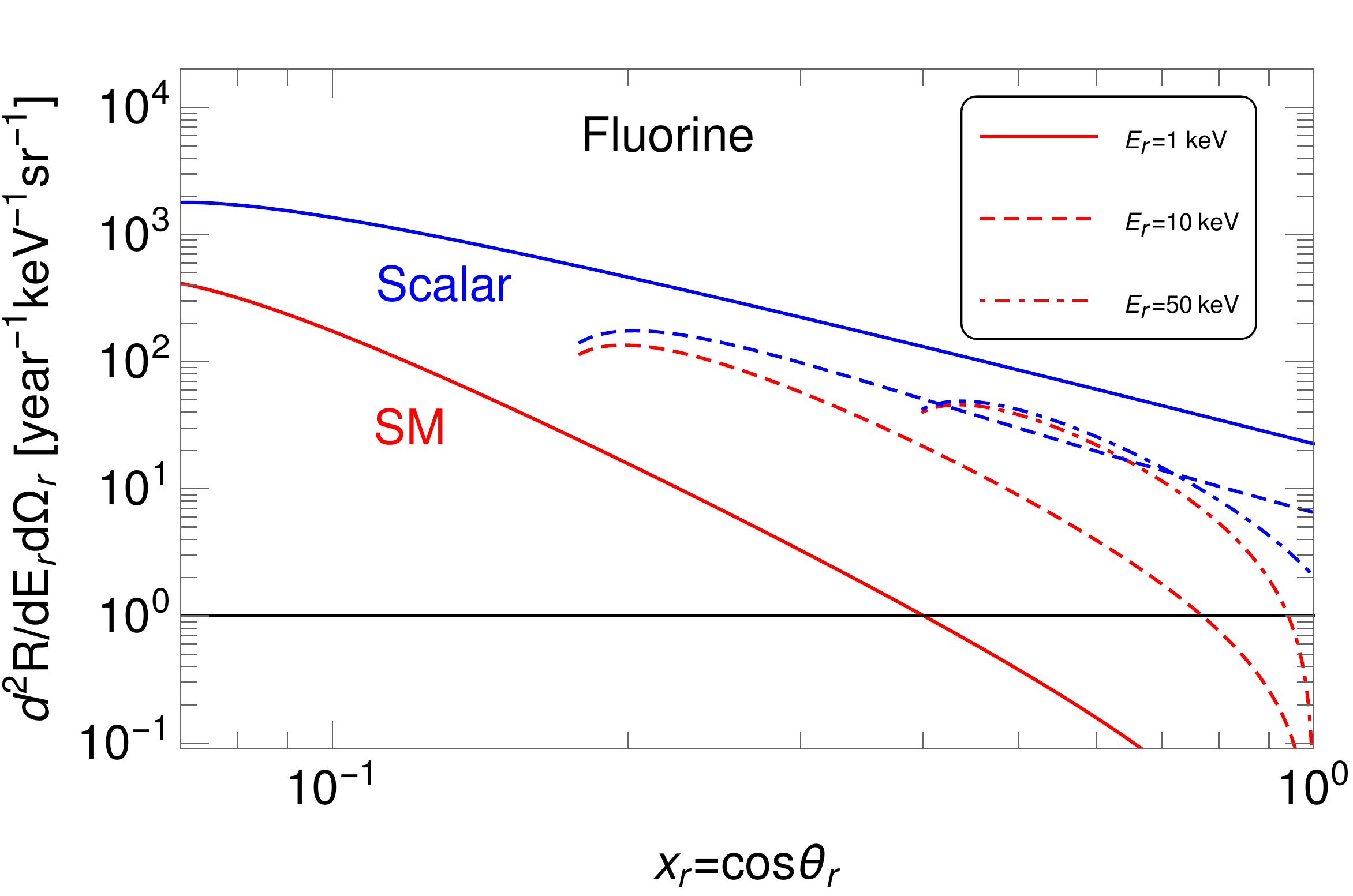}
  \caption{{\bf Left}: DRS slices of fixed recoil energy as a function
    of nuclear recoil scattering angle for a vector mediator (blue
    curve) at a He (Top) and F (bottom) detector using SNS
    neutrinos. The red curve shows the SM result and the black curve
    indicates the single event threshold. {\bf Right}: The same plot
    for the scalar mediator scenario.}
  \label{fig:vector-scalar-momentum-spectra}
\end{figure*}

We will only consider interactions that are lepton flavor universal
and conserving. The vector mediator scenario is described by
\cite{AristizabalSierra:2019ufd,AristizabalSierra:2019ykk}
\begin{equation}
  \label{eq:vec-Lag}
  \mathcal{L}_V=\overline{\nu}\,(f_V+i\gamma_5f_A)\,\gamma_\mu\,\nu\, V^\mu
  +
  \sum_{q=u,d} h_V^q\,\overline{q}\,\gamma_\mu\, q\,V^\mu 
  + 
  \text{H.c.}\ .
\end{equation}
One could also introduce a dark charge leading to CP-violating effects
as done in Ref. \cite{AristizabalSierra:2019ufd}. We do not pursue
such features in this study.

For scalar interactions the set of couplings depends on whether or not
right-handed neutrinos are present. The Lagrangian we use is given by
\cite{Farzan:2018gtr,AristizabalSierra:2019ufd,AristizabalSierra:2019ykk}
\begin{equation}
  \label{eq:sca-Lag}
  \mathcal{L}_\text{S}=\overline{\nu}\,(f_S+i\gamma_5f_S)\,\nu\,S
  +
  \sum_{q=u,d}h_S^q\,\overline{q}\,q\,S
  +
  \text{H.c.}\ .
\end{equation}
In the lepton number violating case the neutrino coupling has to be
recast according to $\nu^T C\,(f_S+i\gamma_5f_S)\,\nu\,S$. As with the
vector mediator, the scalar can be charged under a dark symmetry. We
do not consider axial or pseudoscalar quark couplings since their
contribution to the \cvn\, cross section is small.

The quark-quark operators in Eqs. (\ref{eq:vec-Lag}) and
(\ref{eq:sca-Lag}) induce the following nucleus-nucleus couplings
\begin{align}
  \label{eq:nuc-nuc}
  \text{Vector:}\quad&C_V^N=Z(2h_V^u+h_V^d)+N(h_V^u+2h_V^d)\ ,
                    \nonumber\\
  \text{Scalar:}\quad&C_S^N=Z\sum_qh_q^S\frac{m_n}{m_q}f^n_{T_q}+N\sum_qh_q^S\frac{m_p}{m_q}f^p_{T_q}\ ,
\end{align}
where $m_{n,p}$ are the neutron and proton masses respectively, $q$ is
a quark label, and $f_{T_q}^{n,p}$ refer to hadronic form factors
obtained in chiral perturbation theory using measurements of the
$\pi$-nucleon sigma term
\cite{Crivellin:2013ipa,Hoferichter:2015dsa,Ellis:2000ds}, with the
most up-to-date values given by \cite{Hoferichter:2015dsa}
\begin{alignat}{2}
  \label{eq:ftqn}
  f_{T_u}^p&=(20.8\pm 1.5)\times 10^{-3}\ ,\;&
  f_{T_d}^p=(41.1\pm 2.8)\times 10^{-3}\ ,
  \nonumber\\
  f_{T_u}^n&=(18.9\pm 1.4)\times 10^{-3}\ ,\;&
  f_{T_d}^n=(45.1\pm 2.7)\times 10^{-3}\ .
\end{alignat}
For vector interactions the contributions to the CE$\nu$NS cross
section are obtained from Eq. (\ref{eq:x-sec}) by the substitution
$g_V\to g_V+\xi_V$ \cite{Liao:2017uzy,AristizabalSierra:2019ufd},
where $\xi_V$ reads
\begin{equation}
  \label{eq:xiV}
  \xi_V=\frac{C_V^NF_V}{\sqrt{2}G_F(2m_NE_r+m_V^2)}\ ,
\end{equation}
with $F_V=f_V-if_A$. The combination $g_V+\xi_V$ leads to constructive
or destructive interference depending on the relative sign and size of
the SM and NP contribution. Scalar interactions do not interfere with
the SM at leading order and their contribution to the cross section,
which has to be added to the SM piece Eq. (\ref{eq:x-sec}), is written
as~\cite{Farzan:2018gtr}
\begin{equation}
  \label{eq:xsec-sca}
  \frac{d\sigma_S}{dE_r}=\frac{G_F^2}{2\pi}m_N\xi_S^2
  \frac{m_NE_r}{2E_\nu^2}\ ,
\end{equation}
with the new physics parameters encoded in
\begin{equation}
  \label{eq:xiS}
  \xi_S=\frac{C_S^NF_S}{G_F(2m_NE_r+m_S^2)}\ ,
\end{equation}
where $F_S=f_S-if_P$.

\begin{figure*}
  \centering
  \includegraphics[scale=0.37]{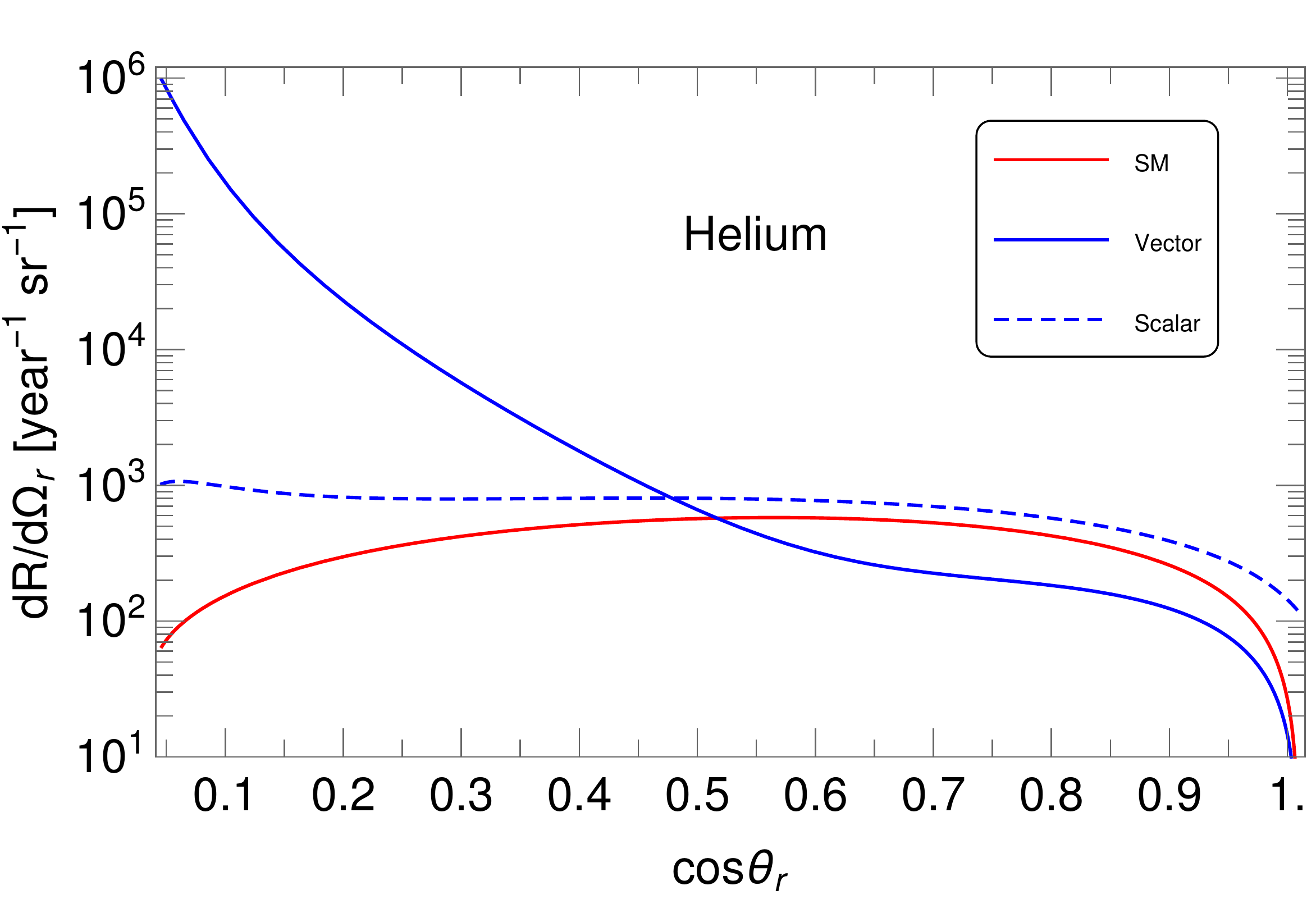}
  \hfill
  \includegraphics[scale=0.37]{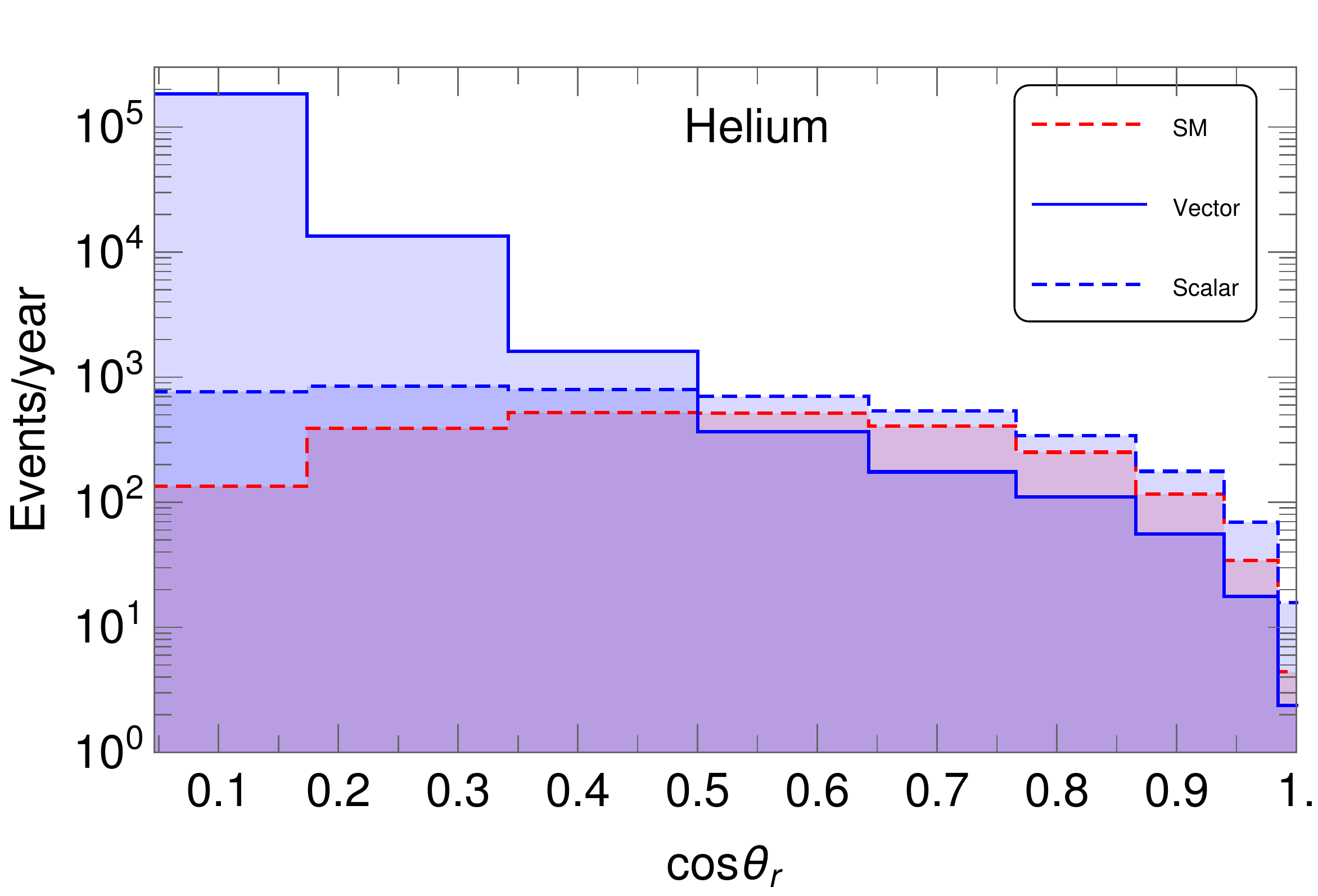}
  \includegraphics[scale=0.37]{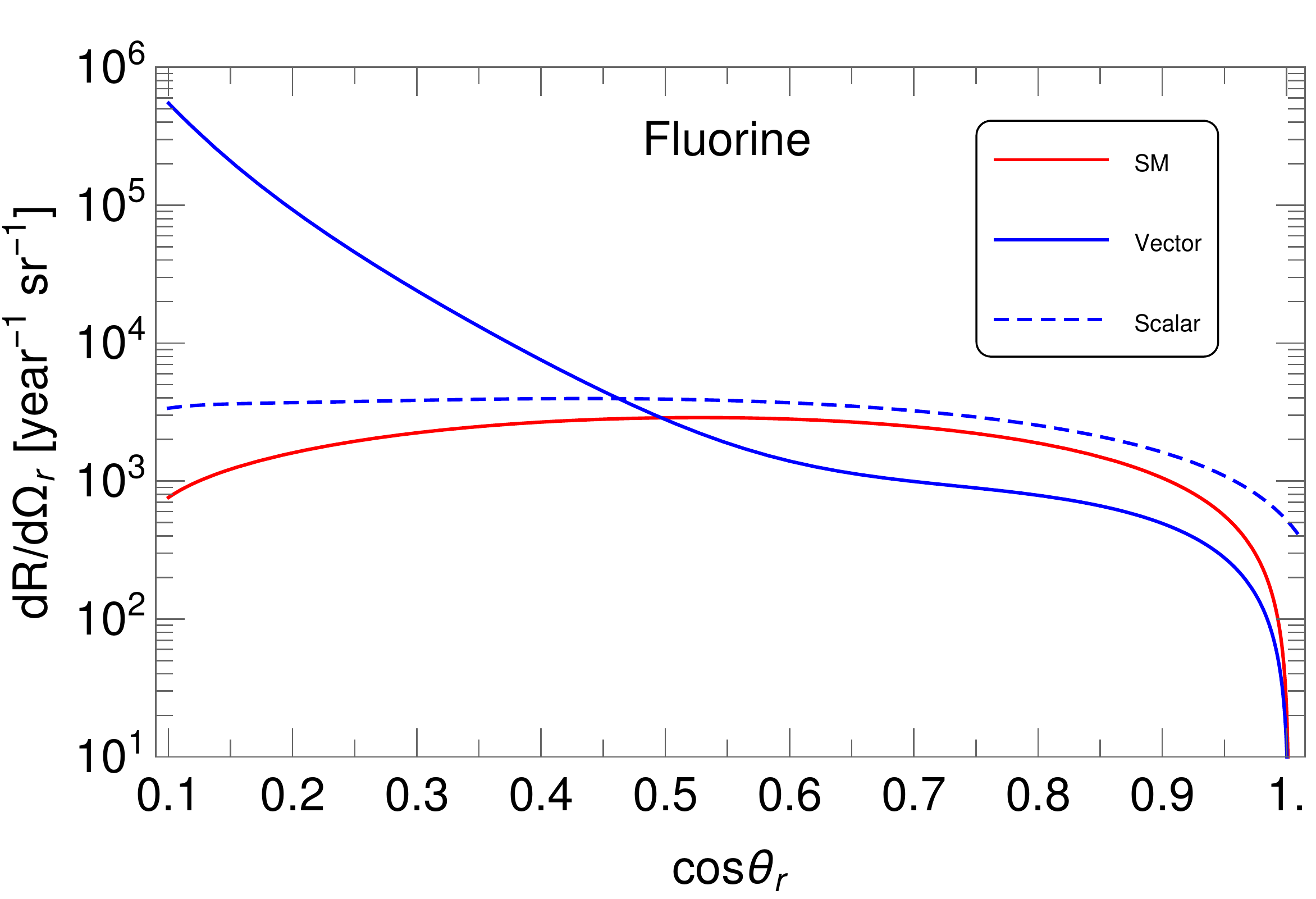}
  \hfill
  \includegraphics[scale=0.37]{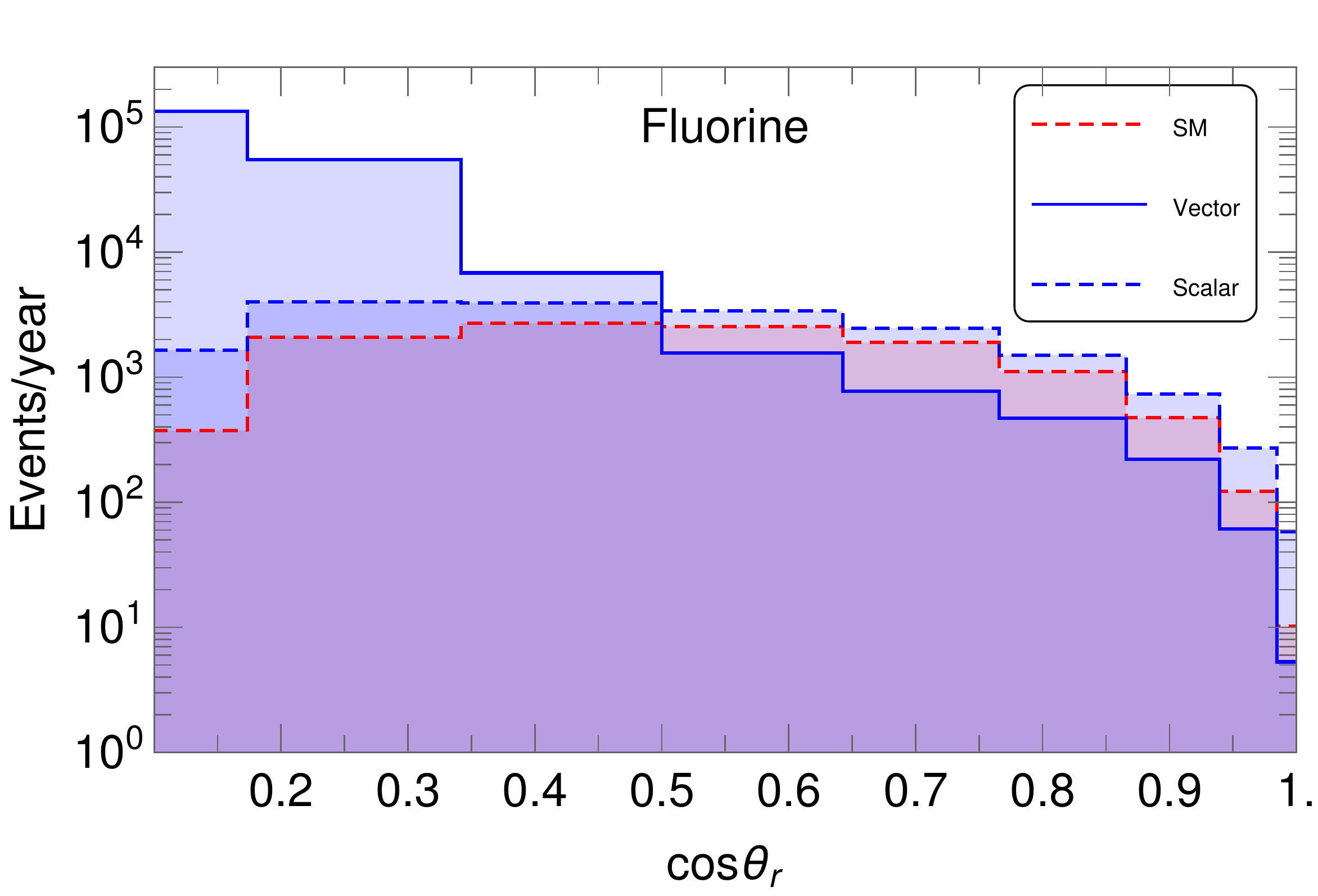}
  \caption{{\bf Left}: The angular distributions in the SM (solid
    red), vector (solid blue) and scalar (dashed blue) for He (Top) and
    F (Bottom) detectors using SNS neutrinos. {\bf Right}: The
    corresponding event yield in angular bins of size $10^\circ$.}
  \label{fig:angular-and-event-spectra-new-interactions}
\end{figure*}
The type of vector and scalar light mediator scenarios described by
the interactions in (\ref{eq:vec-Lag}) and (\ref{eq:sca-Lag}) are
subject to a set of constraints, which have been discussed at length,
for example, in
Refs. \cite{Abdullah:2018ykz,Farzan:2018gtr,AristizabalSierra:2019ufd,AristizabalSierra:2019ykk}. They
can be classified into laboratory bounds, and astrophysical and
cosmological bounds. In the first category most of the limits apply
provided the mediators couple to charged leptons. In our case these
couplings are only present at the one-loop order and so can be safely
ignored. Other limits apply only on the neutrino-quark (nucleon level)
couplings, so they can be readily satisfied without drastically
diminishing the \cvn\, signals. Bounds in the second category can be
tight but are subject to relatively large uncertainties and can be
circumvented through additional new
physics~\cite{Nelson:2007yq,Nelson:2008tn} (an exception are limits
from BBN, see discussion in Sec. \ref{sec:bsm-signals-sns-nus}).

One of the most relevant bounds on the interactions in (\ref{eq:vec-Lag}) and
(\ref{eq:sca-Lag}) comes from COHERENT measurements. A recent study,
using a likelihood analysis that combines energy and timing data,
places bounds for $m_X=1.0\,\,$MeV ($X=V,S$)
\cite{AristizabalSierra:2019ykk}. The bounds are derived using a CsI
target and can be rescaled by $\text{A}_\text{i}/\text{A}_\text{Cs}$
to convert them to the cases of He and F. The resulting bounds are:
\begin{alignat}{2}
\label{eq:coherent-limit}
  \text{He}:&\quad F_VC_V^N&\leq 2.2\times 10^{-8}\ ,\quad 
  F_SC_S^N\leq 1.5\times 10^{-8}\ ,
  \\
  \text{F}:&\quad F_VC_V^N&\leq 1.1\times 10^{-7}\ ,\quad 
  F_SC_S^N\leq 7.3\times 10^{-8}\ .
\end{alignat}
These values generate the maximum number of events consistent with
available data and will be used for the following analysis.
% ------------
% Section
% ------------
\subsection{New physics signals from SNS neutrinos}
\label{sec:bsm-signals-sns-nus}
We can now use Eqs. (\ref{eq:x-sec}), (\ref{eq:momentum-spectrum-3}),
(\ref{eq:xiV}) and (\ref{eq:xsec-sca}) combined with
$g_V\to g_V+\xi_V$ to calculate the DRS in the presence of light
vector and scalar mediators. The results are displayed in
Fig. \ref{fig:vector-scalar-momentum-spectra} for both He and F.

%\begin{figure*}
%  \centering
%  \includegraphics[scale=0.56]{plots/dRdx_mV_He.pdf}
%  \hfill
%  \includegraphics[scale=0.56]{plots/dRdx_mV_F.pdf}
%  \caption{Angular spectrum as a function of \cosr\, in He (left) and
%    F (right) detectors for the SM (red) and with 1, 10, 30, and 50
%    MeV vector mediator masses.}
%  \label{fig:ang-spectrum-mV}
%\end{figure*}

\begin{figure*}
  \centering
  \includegraphics[scale=0.70]{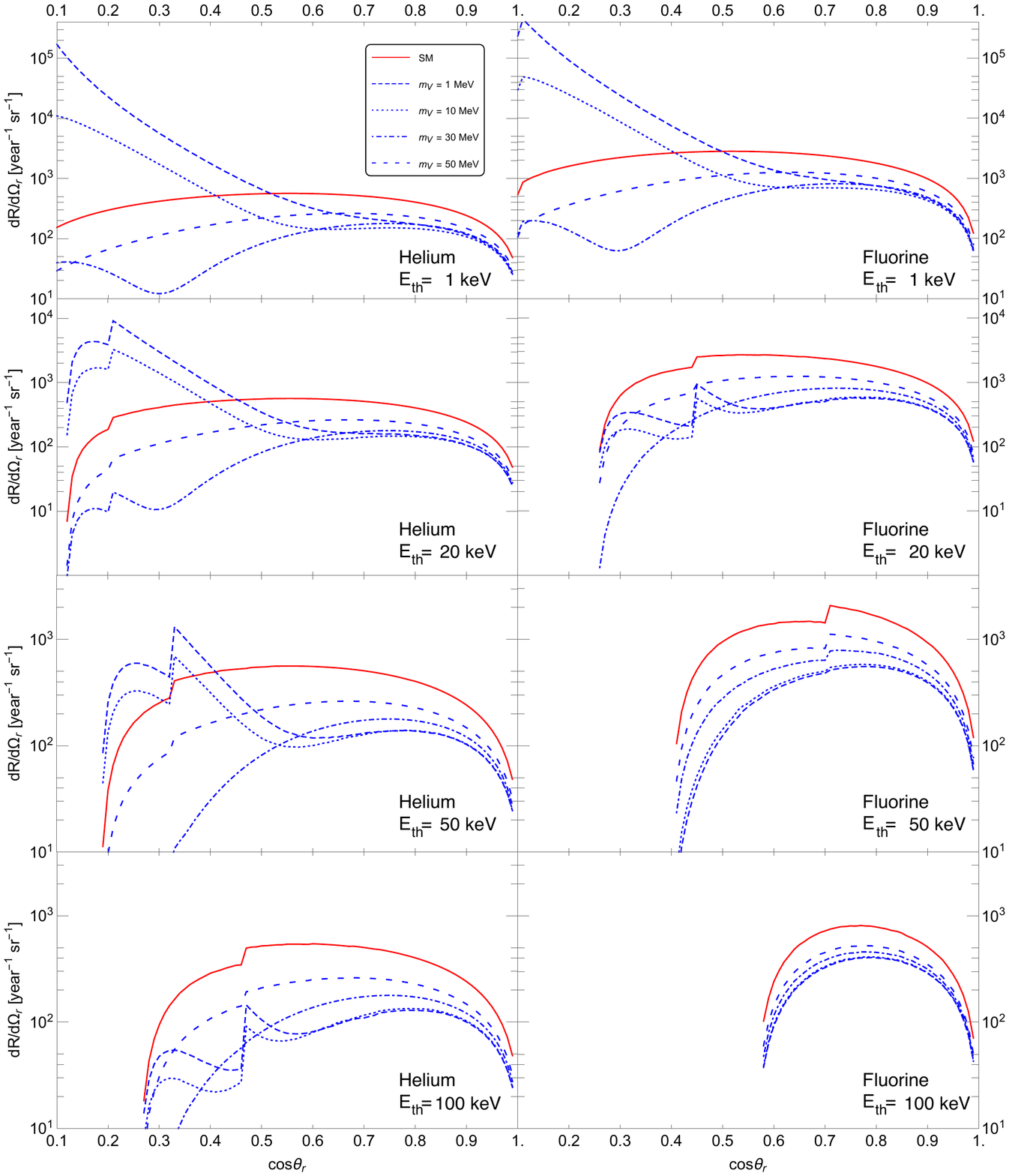}
  \caption{Angular spectrum as a function of \cosr\, in He (left) and
    F (right) detectors for the SM (red) and with 1, 10, 30, and 50
    MeV vector mediator masses. The nuclear recoil energy threshold for
    each panel is indicated.}
  \label{fig:ang-spectrum-mV}
\end{figure*}

For the He case with a vector mediator, all the curves displayed
exhibit a large enhancement bringing them above the single event line
for most of the \cosr\, domain. As we will show there is not always an
enhancement, and in the case of F the $E_r=50$ keV curve with the
presence of a vector is far below the SM analogue due to destructive
interference. The behavior near \cosr$=1$ is mostly unchanged since
the SM cross section is also vector mediated. For the scalar case, we
can see that the enhancement is larger in the forward
direction. However, the enhancement over the SM is significantly
smaller than in the vector case even with the He detector.

\begin{figure*}
  \centering
  \includegraphics[scale=0.37]{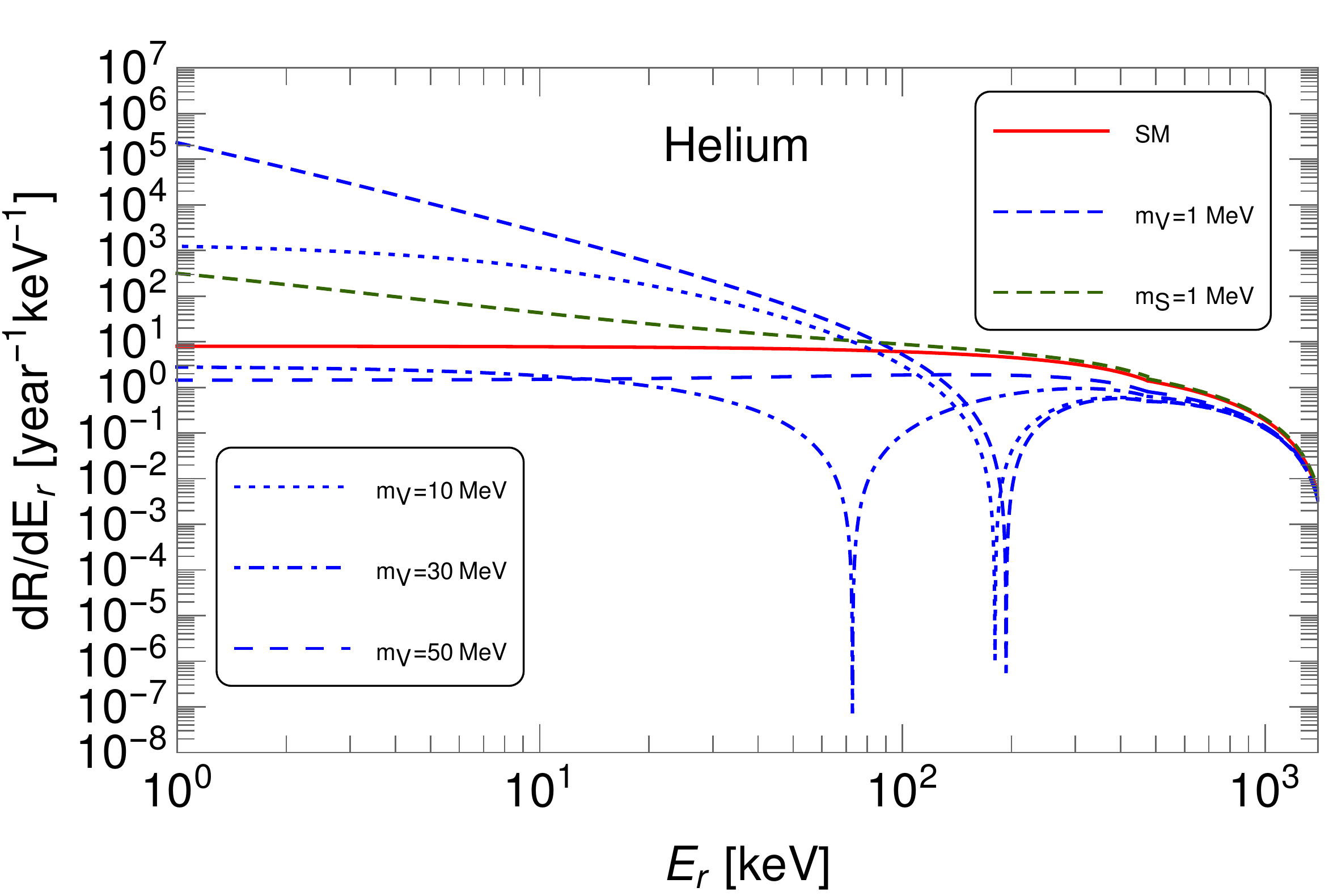}
  \includegraphics[scale=0.37]{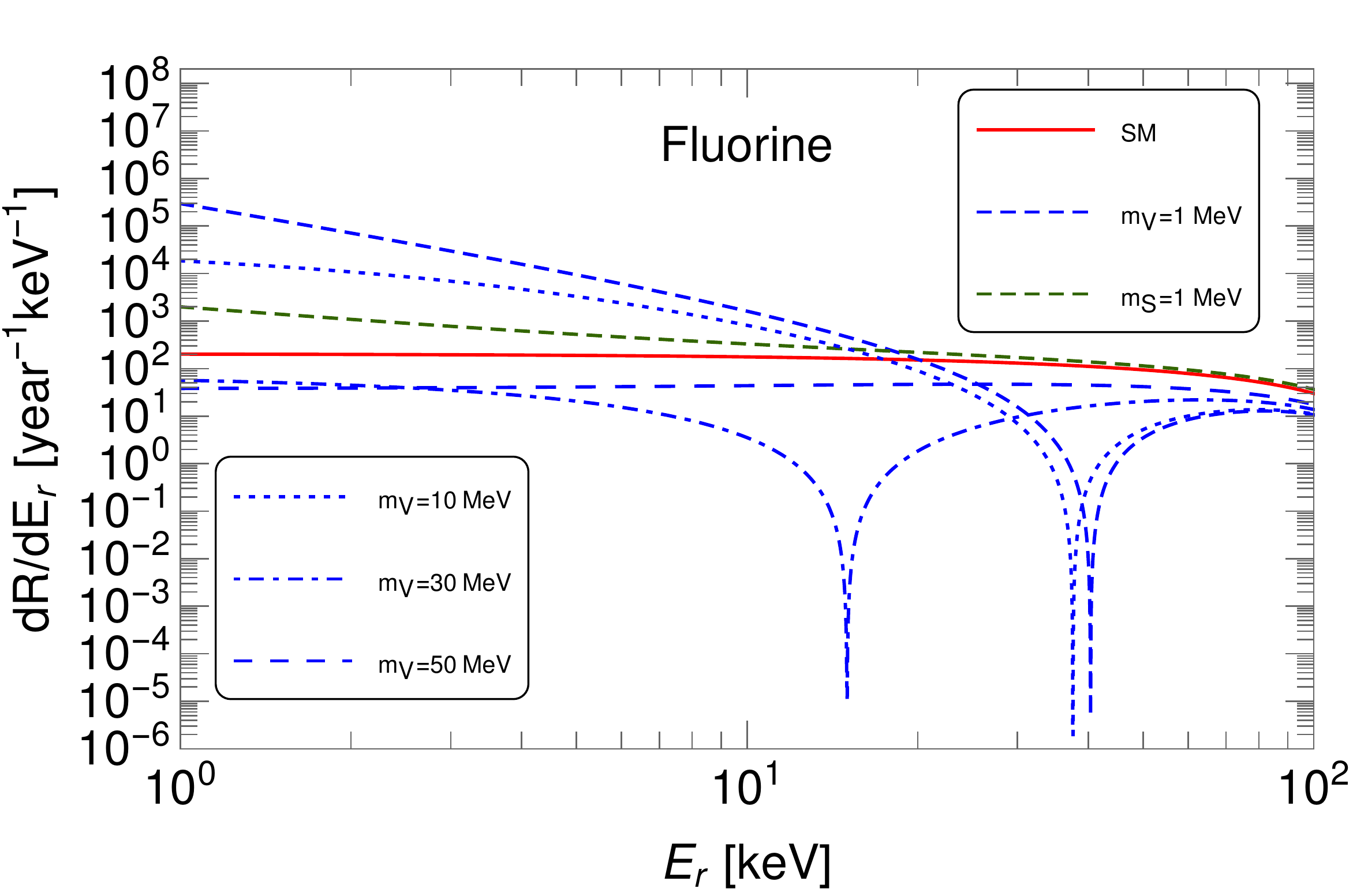}
  \caption{The recoil energy spectrum in the SM (solid red), vector
    (solid blue) and scalar (dashed blue) for He (Left) and F (Right)
    detectors using SNS neutrinos. Included are recoil spectra for
    different vector boson masses.}
  \label{fig:recoil-spectrum}
\end{figure*}
The angular spectrum is shown in
Fig. \ref{fig:angular-and-event-spectra-new-interactions} where the NP
features can be seen more transparently. The vector leads to a modest
deficit for \cosr$>0.5$ while the scalar leads to a small
enhancement. For \cosr$<0.5$ the rate grows tremendously as we
approach \cosr$=0$ in the vector scenario. In contrast, the rate
remains constant in the scalar scenario but with a sizable excess over
the SM at a He detector.

So far we have discussed the results for $1\,\text{MeV}\;$ mediator
masses. However, such species suffer from the tight constraints on the
number of effective relativistic degrees of freedom in the early
universe. This is encoded in the quantity $\Delta N_\text{eff}$ which is precisely determined
through the CMB measurements by the Planck satellite
experiment~\cite{Aghanim:2018eyx}. The constraint has been considered in the
context of light mediator models recently
~\cite{Kamada:2015era,Knapen:2017xzo,Escudero:2019gzq,Sabti:2019mhn,
  Dutta:2020jsy}.  The light mediator contribution to
$\Delta N_\text{eff}$ can be made negligible while contributing
significantly to CE$\nu$NS by making the mediator heavy enough (larger
than a few MeV) that its abundance is negligible due to Boltzmann
suppression at the time of neutrino decoupling. Below, we 
discuss scenarios with mediator masses $>$1 MeV.

To examine how the shape distortion changes with the mediator mass, we
plot the angular spectrum as a function of \cosr\, for mediators
masses 1, 10, 30, and 50 MeV in Fig. \ref{fig:ang-spectrum-mV}, for different
assumed recoil threshold energies. The discontintuities occur due to prompt
neutrinos being unable to induce recoils above a certain angle for
a given detection energy threshold (the analogue of Eq. (\ref{eq:lower-bound-cos}) for
prompt neutrino energies).

For
He, a 30 MeV mediator still modifies the shape of the distribution
although with a deficit instead of an excess, while a 50 MeV mediator
only leads to a rescaling of the SM spectrum. For F, at 30 MeV
mediator mass the NP spectrum is already a rescaling. As the energy threshold is
increased, both the discontinuity and the lower end of the distribution move
towards larger \cosr.

Since detectors sensitive to $E_r$ already exist and angular
information could come at an expense of energy resolution, it is
useful to compare the angular spectrum with the associated energy
spectrum (Fig. \ref{fig:recoil-spectrum}). The vector induced deficit
is more dramatic than in the angular distribution and occurs at large
values of $E_r$, which are accessible with current technology. The
scalar curve at high recoil energies coincides with that of the
SM. Finally, one can note that dips in the recoil spectra are smeared
out in nuclear angle space. For example, in helium and for
$m_V=30\,$MeV the recoil spectra exhibits a well localized dip at
about $E_r=70\,$keV. At the angular distribution level, that sharp
downward spike results in a way less pronounced feature at
$\cos\theta_r\simeq 0.3$.

It is insightful to use Fig. \ref{fig:recoil-spectrum} in conjunction with
Fig. \ref{fig:ang-spectrum-mV} to understand the effect of detector thresholds
on observables. From Fig. \ref{fig:recoil-spectrum} one can directly read off the recoil spectrum from any 
energy threshold between 1 keV and 100 keV, and a higher value necessarily 
leads to lower NP sensitivity. This can be compared to one of the representative
threshold values in Fig. \ref{fig:ang-spectrum-mV} to see how the shape discrimination
appears in the angle domain.

The excess regions are more interesting to compare since with larger
signals backgrounds and systematic errors become less
challenging. Comparing the plots we can see a qualitatively unique feature
in the lower~\cosr~distribution compared to that of low $E_r$: the three scenarios (SM, SM+Vector, and SM+Scalar) lead to slopes that are negative, positive and vanishing respectively.
 The discriminating region in the energy domain is roughly between 1 keV
and 100 keV, while in the angle domain it is between 85$^\circ$ and
60$^\circ$. It is unclear at this stage which choice would lead to
stronger limits. For that a likelihood analysis using various
combinations of detector resolutions is necessary.

One crucial difference is that increasing the detection threshold,
say, to 10 keV would eliminate a large portion of the signal
discrimination region, while the small \cosr\, region would still be
accessible. In other words, it could be beneficial to trade a higher
detection threshold with finer angular resolution at large angles. The
ratios of events with NP to that in the SM are
\begin{align}
  \text{He:} &&N_\text{V}/N_\text{SM} &=& 106 &,&  N_\text{S}/N_\text{SM} &=& 1.8\ , \\
  \text{F:} &&N_\text{V}/N_\text{SM} &=& 23 &,&  N_\text{S}/N_\text{SM} &=& 1.6\ .
\end{align}

\begin{figure*}
  \centering
  \includegraphics[scale=0.37]{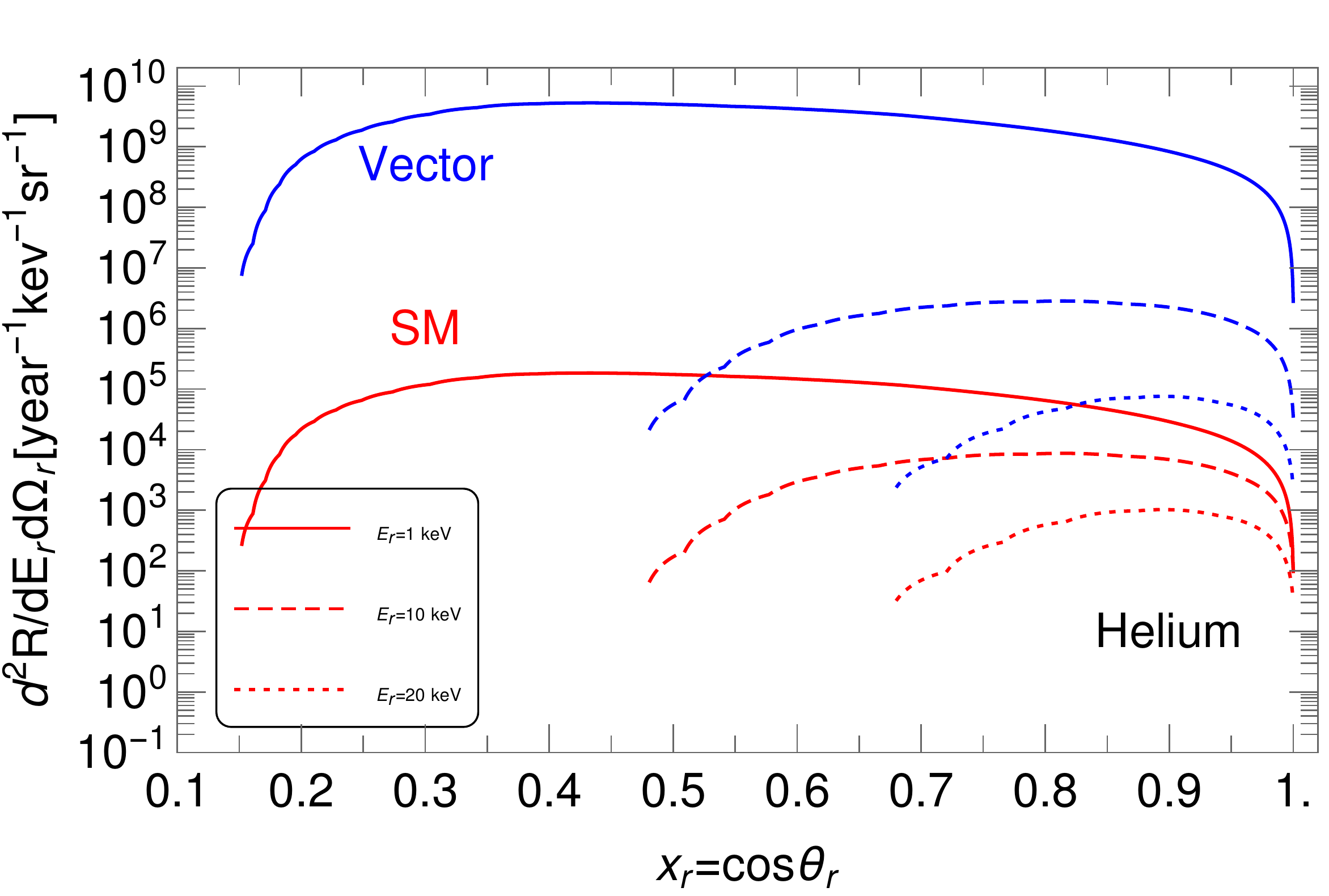}
  \hfill
  \includegraphics[scale=0.37]{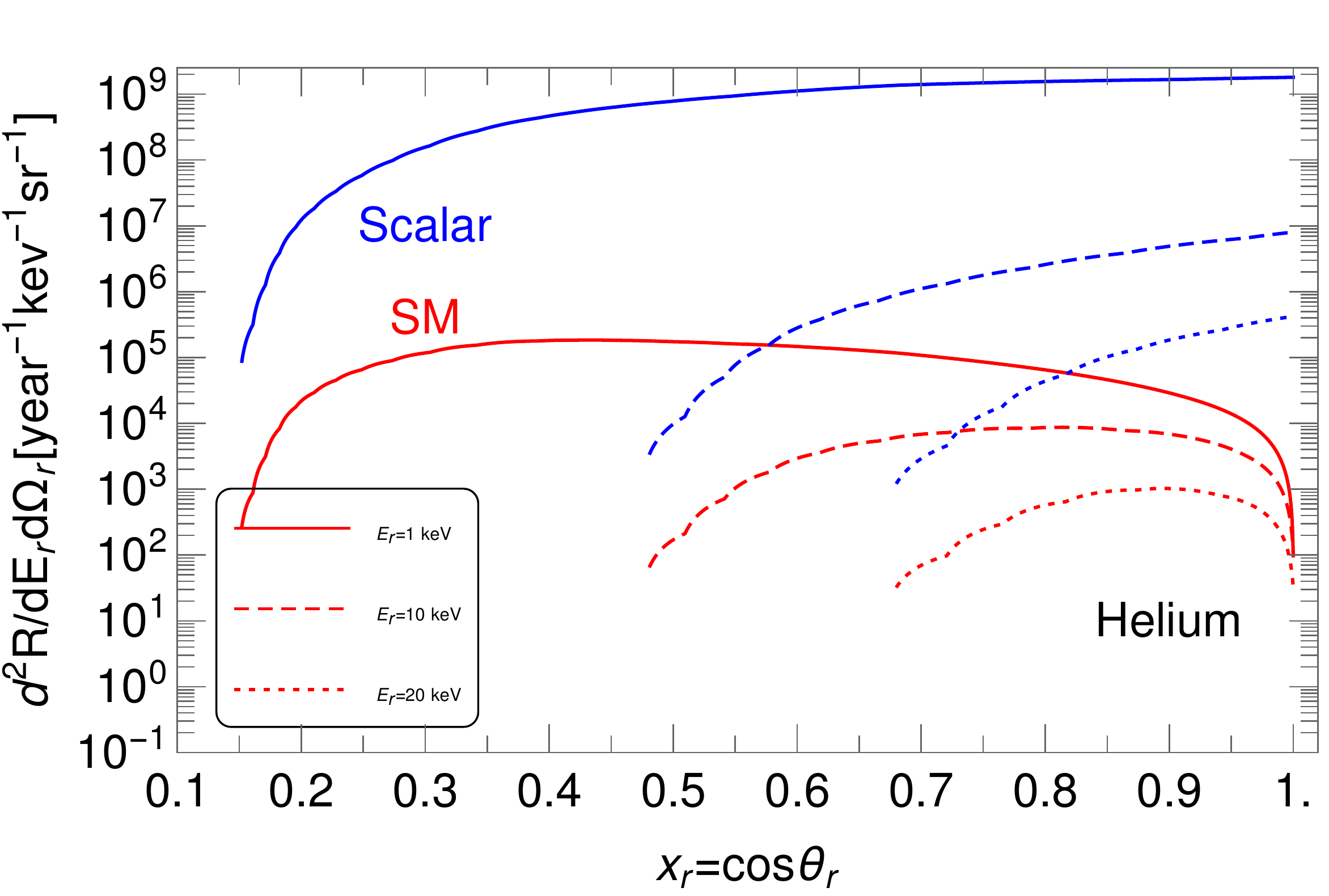}
  \includegraphics[scale=0.37]{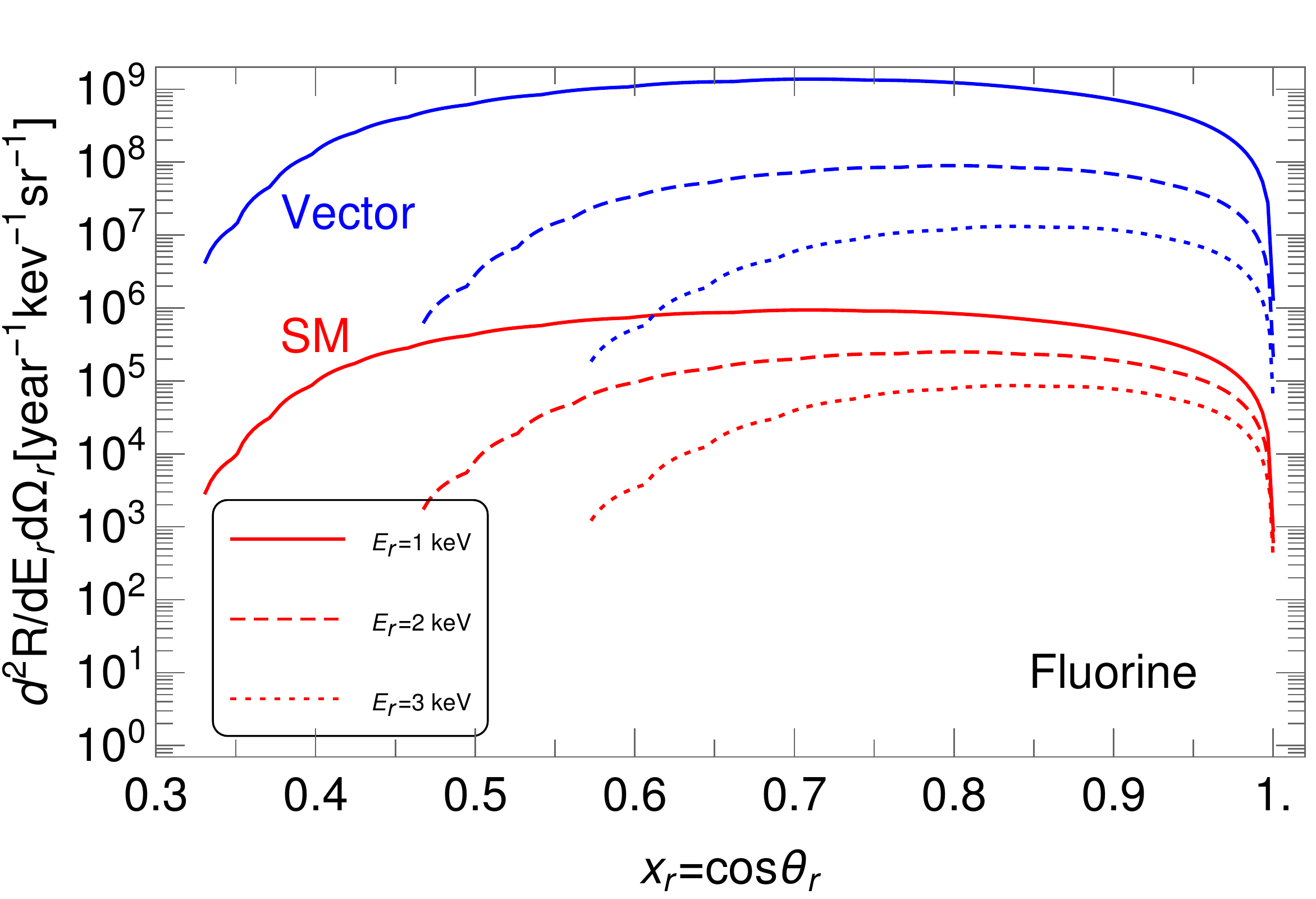}
  \hfill
  \includegraphics[scale=0.37]{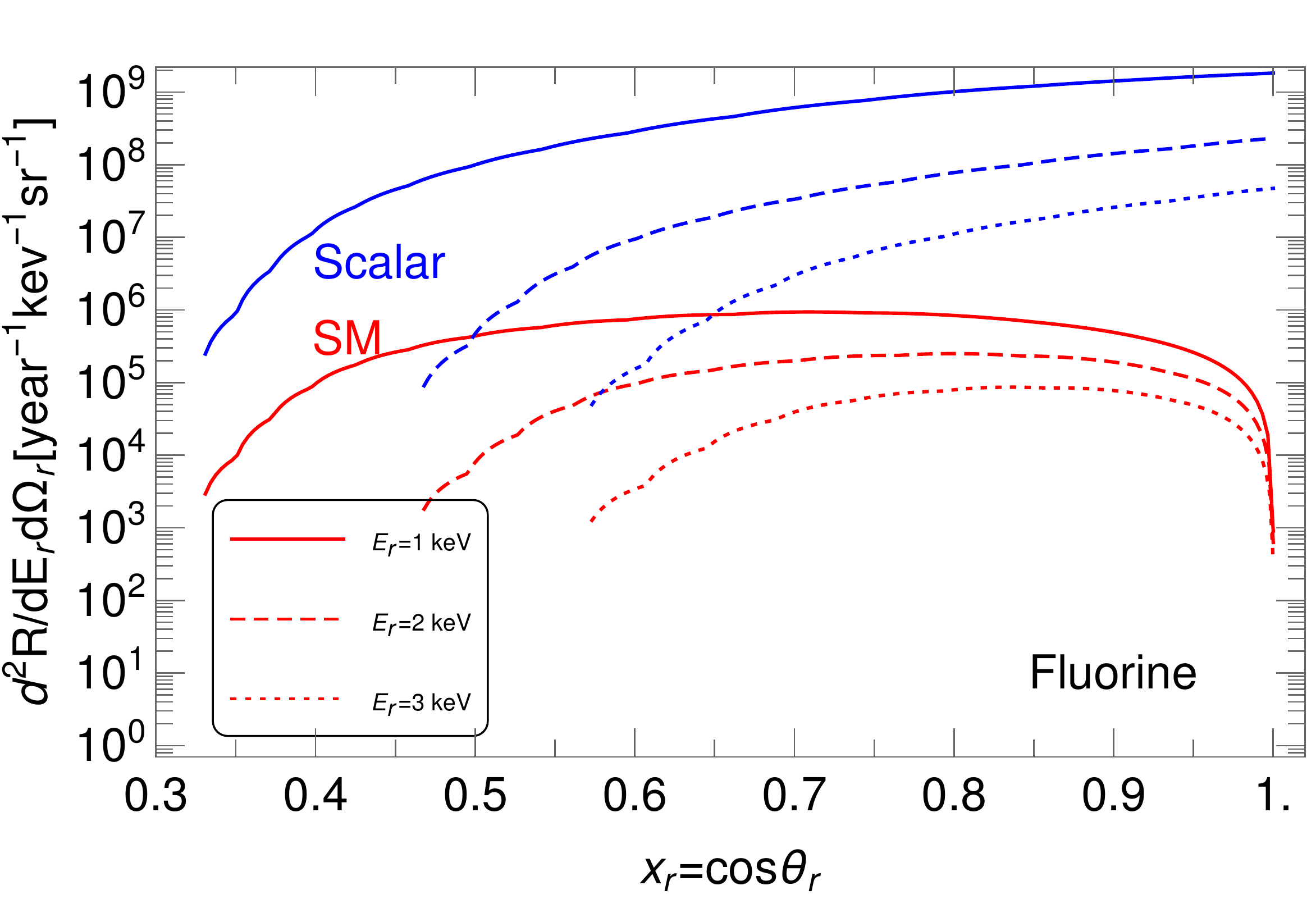}
  \caption{{\bf Left}: DRS slices of fixed recoil energy as a function
    of nuclear recoil scattering angle for a vector mediator (blue
    curve) at an He (Top) and F (bottom) detector using reactor
    neutrinos. The red curve shows the SM result. {\bf Right}: The
    same plot for the scalar mediator scenario.}
  \label{fig:vector-scalar-momentum-spectra-reactor}
\end{figure*}

\begin{figure*}
  \centering
  \includegraphics[scale=0.37]{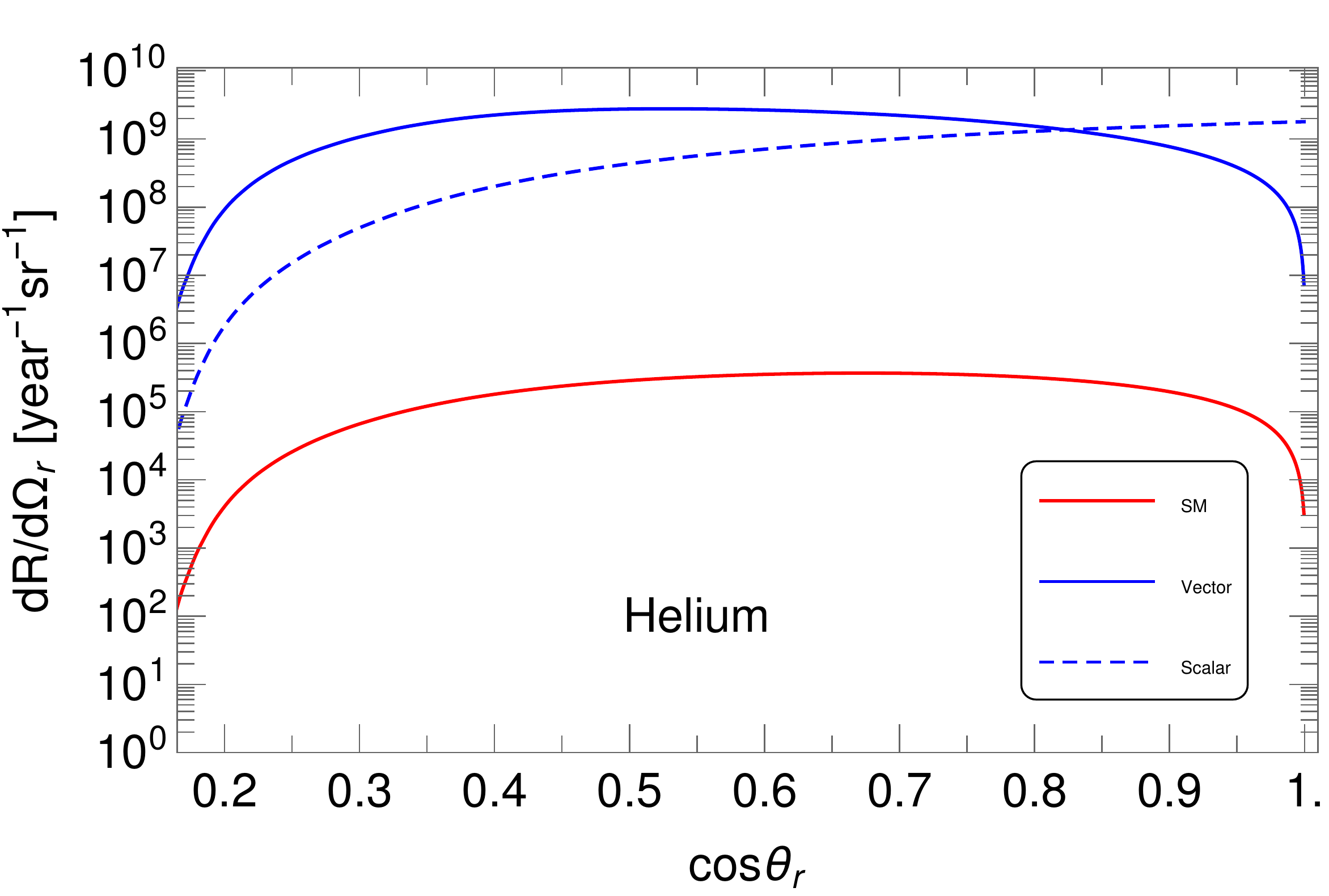}
  \hfill
  \includegraphics[scale=0.37]{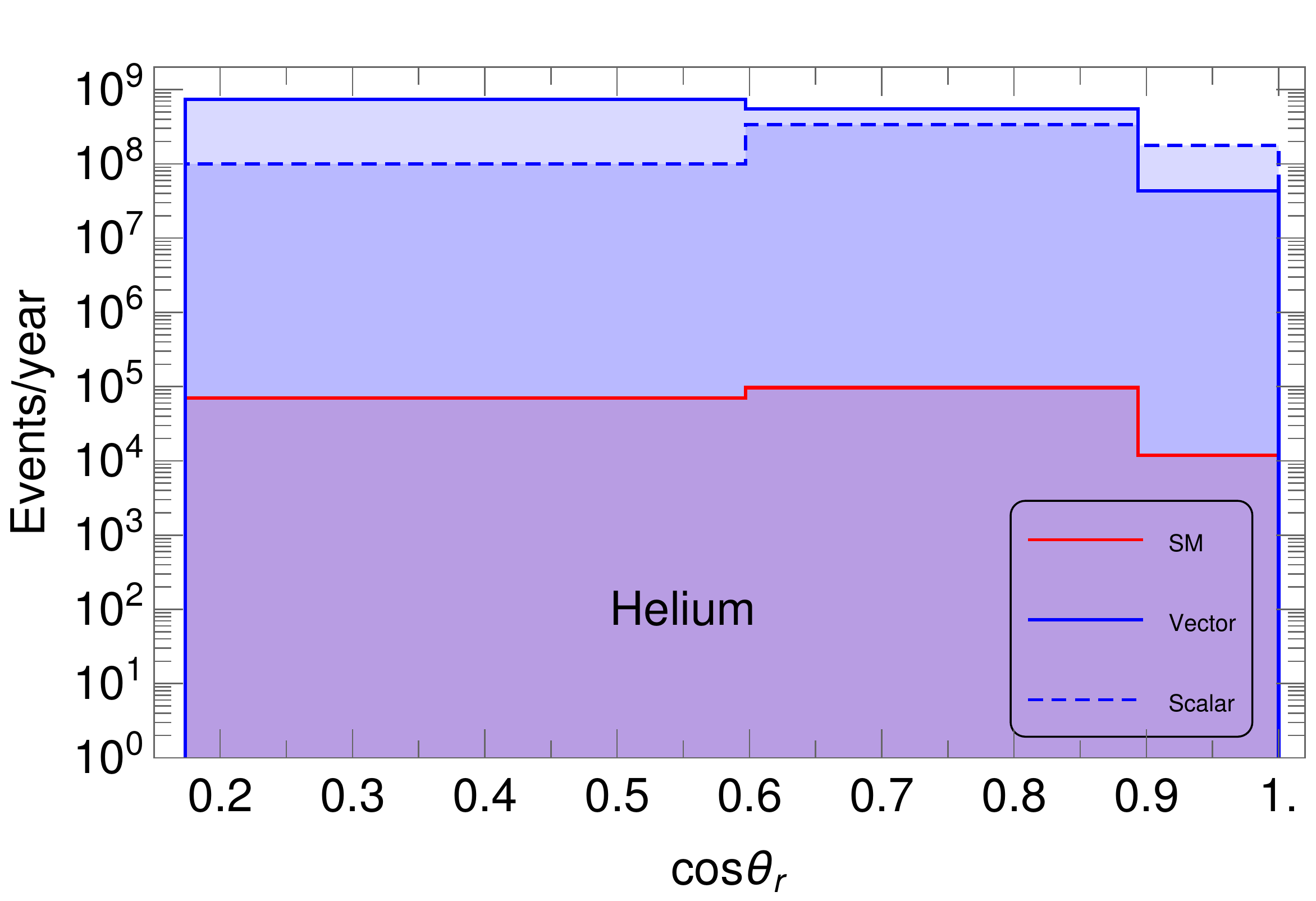}
  \includegraphics[scale=0.37]{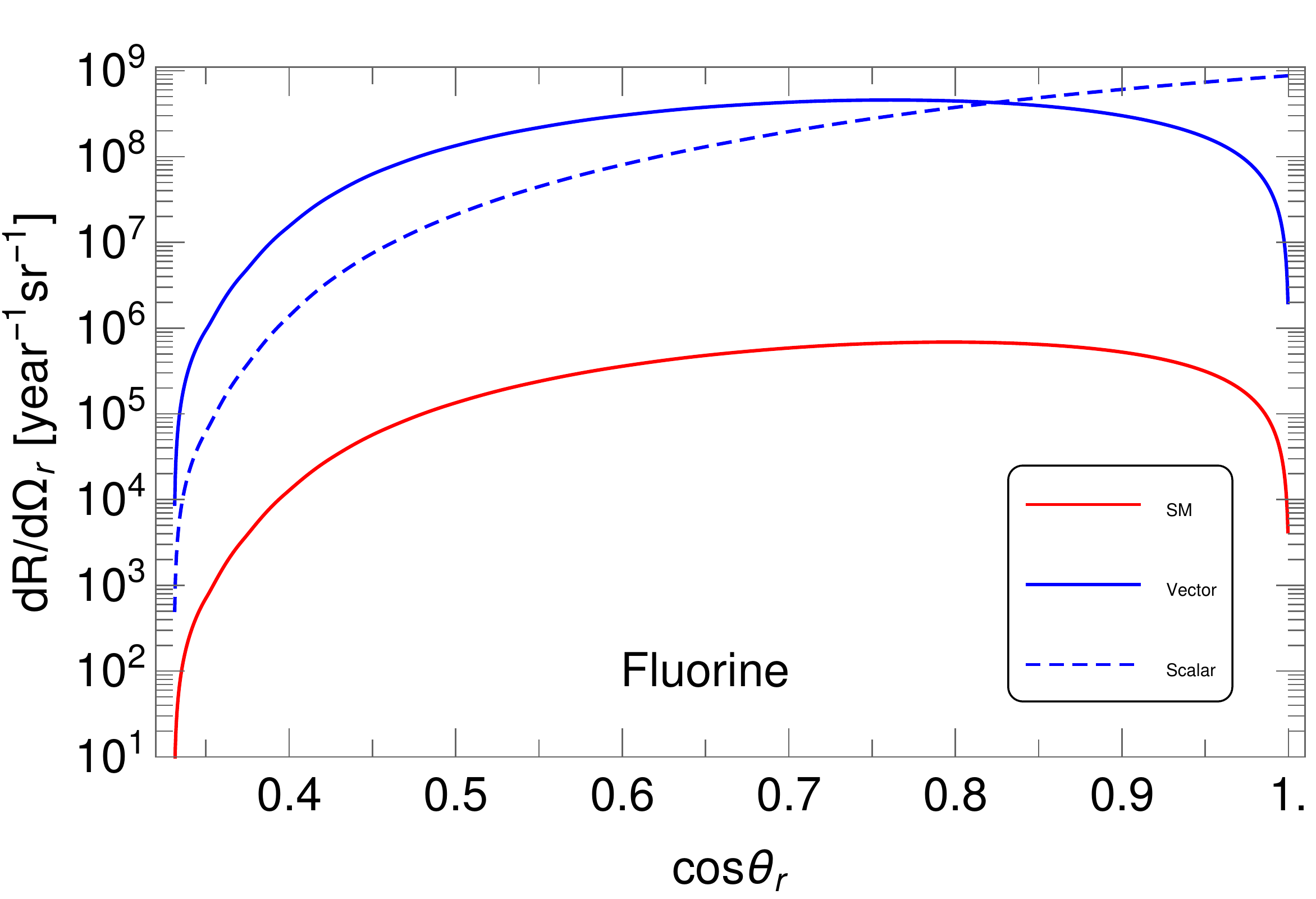}
  \hfill
  \includegraphics[scale=0.37]{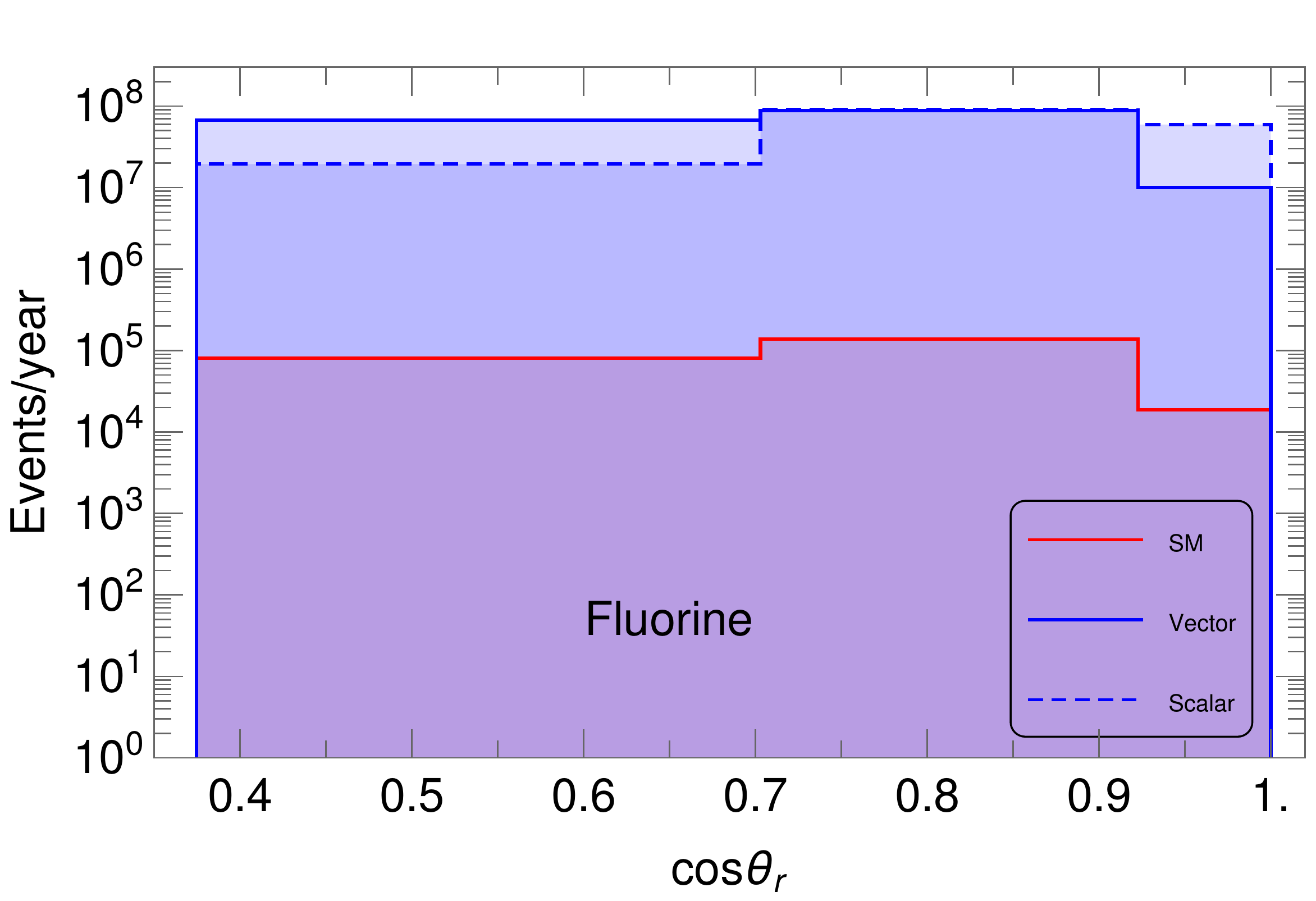}
  \caption{{\bf Left}: The angular distributions in the SM (solid
    red), vector (solid blue) and scalar (dashed blue) for He (Top) and
    F (Bottom) detectors using reactor neutrinos. {\bf Right}: The
    corresponding event yield in angular bins of size $30^\circ$.}
  \label{fig:angular-and-event-spectra-new-interactions-reactor}
\end{figure*}
% --------------
% Section
% --------------
\subsection{New physics signals from reactor neutrinos}
\label{sec:bsm-signals-reactor-nus}
Finally, we turn to new physics signals at reactors. The event rate is
enormous, though as is seen in the DRS slices in
Fig. \ref{fig:vector-scalar-momentum-spectra-reactor}, most of the
events have very low $E_r$, particularly in the F case. Examining the
angular distribution in
Fig. \ref{fig:angular-and-event-spectra-new-interactions-reactor}, we
see that the distribution in the vector scenario is similar to that of
the SM and differs only by a scaling factor. This is due in part to
the vector nature of the SM interaction and in part due to energy
scale of the recoil being much smaller than the mass of the mediator
(1 MeV). With an SNS like source, the differences in shapes persist
even with 10 MeV mediators but cease at values closer to 100 MeV.

In contrast, the scalar mediator leads to a qualitatively different
spectral shape which could potentially be resolved with enough data,
and unlike in the SNS source, the excess over the SM is substantial.
\section{Conclusions}
\label{sec:conclusions}%
We have performed a theoretical study of the directional behavior of
\cvn\, using stopped-pion and reactor sources. We consider gaseous
helium and fluorine detectors, and generate predictions for the SM
nuclear recoil distributions. In addition, we consider scenarios with
the addition of light vector or scalar mediators. These light
mediators can arise in the context of anomaly free $U(1)_{B-L}$, $U(1)_{T_{3R}}$, $U(1)_{L_\mu-L_\tau}$
symmetry models. In the context of new symmetry models, we can have multiple copies of same type of mediators and/or different types of mediators being present. The direction information would provide an important additional handle to investigate all these new models. We have 
identified angular features that can aid in identifying vector
mediators at a stopped-pion source such as SNS, and for scalar
mediators at reactors. We also provided information on the interplay
between energy sensitivity and threshold, and directional sensitivity.

Though our analysis has focused on \cvn\, and how new physics may be
extracted though neutrino interactions, it would also be interesting
to extend our analysis to understand the importance of directionality
in low mass dark matter searches using both stopped-pion and reactor
sources. Stopped-pion based experiments like COHERENT have been shown to be
valuable probes of sub-GeV dark matter~\cite{Akimov:2019xdj},
especially since timing and recoil energy information is able to
effectively reduce SM and experimental
backgrounds~\cite{Dutta:2019nbn,Akimov:2019xdj}. Extending beyond nuclear recoils, it
is also interesting to considering directionality in electron
recoils. This may even provide new means to discriminate backgrounds
and identify new signals via Migdal electrons~\cite{Ibe:2017yqa}. Even
for energy-only based analyses, including the Migdal effect has been shown
to improve bounds on low-mass dark matter in xenon
detectors~\cite{Aprile:2019jmx}.

An obvious next step is to perform a more thorough, likelihood based
analysis using more realistic modeling of the experimental setup. For example, this
includes modeling the source as an extended
object which leads to angular uncertainty. Another example is accounting for backgrounds
which limit the significance of the signal. Using realistic fiducial
detector masses, which would likely be smaller than the values used
here, is necessary for accurate estimates of exposure.
More importantly, factoring in the prospective efficiency curves as well as
the spatial, energy, and angular resolutions could dramatically
alter all the spectral shapes in this study and reframe the interplay
between energy and directional information. Directionality in \cvn\, is a new and unexplored territory with new
ideas, questions and answers to be tapped by the neutrino community.

\section*{Acknowledgements}
We thank Neil Spooner, Sven Vahsen, Kate Scholberg, and Phil Barbeau
discussions on this paper.  DAS is supported by the grant ``Unraveling
new physics in the high-intensity and high-energy frontiers'',
Fondecyt No 1171136.  BD and LES acknowledge support from DOE Grant
de-sc0010813.  We thank the organizers of the ``Magnificent CEvNS 2019
Workshop'' where this work was initiated.

\bibliography{references}

\begin{thebibliography}{60}
\expandafter\ifx\csname natexlab\endcsname\relax\def\natexlab#1{#1}\fi
\expandafter\ifx\csname bibnamefont\endcsname\relax
  \def\bibnamefont#1{#1}\fi
\expandafter\ifx\csname bibfnamefont\endcsname\relax
  \def\bibfnamefont#1{#1}\fi
\expandafter\ifx\csname citenamefont\endcsname\relax
  \def\citenamefont#1{#1}\fi
\expandafter\ifx\csname url\endcsname\relax
  \def\url#1{\texttt{#1}}\fi
\expandafter\ifx\csname urlprefix\endcsname\relax\def\urlprefix{URL }\fi
\providecommand{\bibinfo}[2]{#2}
\providecommand{\eprint}[2][]{\url{#2}}

\bibitem[{\citenamefont{Akimov et~al.}(2017)}]{Akimov:2017ade}
\bibinfo{author}{\bibfnamefont{D.}~\bibnamefont{Akimov}} \bibnamefont{et~al.}
  (\bibinfo{collaboration}{COHERENT}), \bibinfo{journal}{Science}
  (\bibinfo{year}{2017}), \eprint{1708.01294}.

\bibitem[{\citenamefont{Akimov et~al.}(2020)}]{Akimov:2020pdx}
\bibinfo{author}{\bibfnamefont{D.}~\bibnamefont{Akimov}} \bibnamefont{et~al.}
  (\bibinfo{collaboration}{COHERENT}) (\bibinfo{year}{2020}),
  \eprint{2003.10630}.

\bibitem[{\citenamefont{Coloma et~al.}(2017{\natexlab{a}})\citenamefont{Coloma,
  Denton, Gonzalez-Garcia, Maltoni, and Schwetz}}]{Coloma:2017egw}
\bibinfo{author}{\bibfnamefont{P.}~\bibnamefont{Coloma}},
  \bibinfo{author}{\bibfnamefont{P.~B.} \bibnamefont{Denton}},
  \bibinfo{author}{\bibfnamefont{M.~C.} \bibnamefont{Gonzalez-Garcia}},
  \bibinfo{author}{\bibfnamefont{M.}~\bibnamefont{Maltoni}}, \bibnamefont{and}
  \bibinfo{author}{\bibfnamefont{T.}~\bibnamefont{Schwetz}},
  \bibinfo{journal}{JHEP} \textbf{\bibinfo{volume}{04}}, \bibinfo{pages}{116}
  (\bibinfo{year}{2017}{\natexlab{a}}), \eprint{1701.04828}.

\bibitem[{\citenamefont{Coloma et~al.}(2017{\natexlab{b}})\citenamefont{Coloma,
  Gonzalez-Garcia, Maltoni, and Schwetz}}]{Coloma:2017ncl}
\bibinfo{author}{\bibfnamefont{P.}~\bibnamefont{Coloma}},
  \bibinfo{author}{\bibfnamefont{M.~C.} \bibnamefont{Gonzalez-Garcia}},
  \bibinfo{author}{\bibfnamefont{M.}~\bibnamefont{Maltoni}}, \bibnamefont{and}
  \bibinfo{author}{\bibfnamefont{T.}~\bibnamefont{Schwetz}}
  (\bibinfo{year}{2017}{\natexlab{b}}), \eprint{1708.02899}.

\bibitem[{\citenamefont{Liao and Marfatia}(2017)}]{Liao:2017uzy}
\bibinfo{author}{\bibfnamefont{J.}~\bibnamefont{Liao}} \bibnamefont{and}
  \bibinfo{author}{\bibfnamefont{D.}~\bibnamefont{Marfatia}},
  \bibinfo{journal}{Phys. Lett.} \textbf{\bibinfo{volume}{B775}},
  \bibinfo{pages}{54} (\bibinfo{year}{2017}), \eprint{1708.04255}.

\bibitem[{\citenamefont{Dent et~al.}(2018)\citenamefont{Dent, Dutta, Liao,
  Newstead, Strigari, and Walker}}]{Dent:2017mpr}
\bibinfo{author}{\bibfnamefont{J.~B.} \bibnamefont{Dent}},
  \bibinfo{author}{\bibfnamefont{B.}~\bibnamefont{Dutta}},
  \bibinfo{author}{\bibfnamefont{S.}~\bibnamefont{Liao}},
  \bibinfo{author}{\bibfnamefont{J.~L.} \bibnamefont{Newstead}},
  \bibinfo{author}{\bibfnamefont{L.~E.} \bibnamefont{Strigari}},
  \bibnamefont{and} \bibinfo{author}{\bibfnamefont{J.~W.}
  \bibnamefont{Walker}}, \bibinfo{journal}{Phys. Rev.}
  \textbf{\bibinfo{volume}{D97}}, \bibinfo{pages}{035009}
  (\bibinfo{year}{2018}), \eprint{1711.03521}.

\bibitem[{\citenamefont{Papoulias and Kosmas}(2018)}]{Kosmas:2017tsq}
\bibinfo{author}{\bibfnamefont{D.~K.} \bibnamefont{Papoulias}}
  \bibnamefont{and} \bibinfo{author}{\bibfnamefont{T.~S.}
  \bibnamefont{Kosmas}}, \bibinfo{journal}{Phys. Rev.}
  \textbf{\bibinfo{volume}{D97}}, \bibinfo{pages}{033003}
  (\bibinfo{year}{2018}), \eprint{1711.09773}.

\bibitem[{\citenamefont{Billard et~al.}(2018)\citenamefont{Billard, Johnston,
  and Kavanagh}}]{Billard:2018jnl}
\bibinfo{author}{\bibfnamefont{J.}~\bibnamefont{Billard}},
  \bibinfo{author}{\bibfnamefont{J.}~\bibnamefont{Johnston}}, \bibnamefont{and}
  \bibinfo{author}{\bibfnamefont{B.~J.} \bibnamefont{Kavanagh}}
  (\bibinfo{year}{2018}), \eprint{1805.01798}.

\bibitem[{\citenamefont{Lindner et~al.}(2017)\citenamefont{Lindner, Rodejohann,
  and Xu}}]{Lindner:2016wff}
\bibinfo{author}{\bibfnamefont{M.}~\bibnamefont{Lindner}},
  \bibinfo{author}{\bibfnamefont{W.}~\bibnamefont{Rodejohann}},
  \bibnamefont{and} \bibinfo{author}{\bibfnamefont{X.-J.} \bibnamefont{Xu}},
  \bibinfo{journal}{JHEP} \textbf{\bibinfo{volume}{03}}, \bibinfo{pages}{097}
  (\bibinfo{year}{2017}), \eprint{1612.04150}.

\bibitem[{\citenamefont{Abdullah et~al.}(2018)\citenamefont{Abdullah, Dent,
  Dutta, Kane, Liao, and Strigari}}]{Abdullah:2018ykz}
\bibinfo{author}{\bibfnamefont{M.}~\bibnamefont{Abdullah}},
  \bibinfo{author}{\bibfnamefont{J.~B.} \bibnamefont{Dent}},
  \bibinfo{author}{\bibfnamefont{B.}~\bibnamefont{Dutta}},
  \bibinfo{author}{\bibfnamefont{G.~L.} \bibnamefont{Kane}},
  \bibinfo{author}{\bibfnamefont{S.}~\bibnamefont{Liao}}, \bibnamefont{and}
  \bibinfo{author}{\bibfnamefont{L.~E.} \bibnamefont{Strigari}},
  \bibinfo{journal}{Phys. Rev.} \textbf{\bibinfo{volume}{D98}},
  \bibinfo{pages}{015005} (\bibinfo{year}{2018}), \eprint{1803.01224}.

\bibitem[{\citenamefont{Farzan et~al.}(2018)\citenamefont{Farzan, Lindner,
  Rodejohann, and Xu}}]{Farzan:2018gtr}
\bibinfo{author}{\bibfnamefont{Y.}~\bibnamefont{Farzan}},
  \bibinfo{author}{\bibfnamefont{M.}~\bibnamefont{Lindner}},
  \bibinfo{author}{\bibfnamefont{W.}~\bibnamefont{Rodejohann}},
  \bibnamefont{and} \bibinfo{author}{\bibfnamefont{X.-J.} \bibnamefont{Xu}},
  \bibinfo{journal}{JHEP} \textbf{\bibinfo{volume}{05}}, \bibinfo{pages}{066}
  (\bibinfo{year}{2018}), \eprint{1802.05171}.

\bibitem[{\citenamefont{Brdar et~al.}(2018)\citenamefont{Brdar, Rodejohann, and
  Xu}}]{Brdar:2018qqj}
\bibinfo{author}{\bibfnamefont{V.}~\bibnamefont{Brdar}},
  \bibinfo{author}{\bibfnamefont{W.}~\bibnamefont{Rodejohann}},
  \bibnamefont{and} \bibinfo{author}{\bibfnamefont{X.-J.} \bibnamefont{Xu}},
  \bibinfo{journal}{JHEP} \textbf{\bibinfo{volume}{12}}, \bibinfo{pages}{024}
  (\bibinfo{year}{2018}), \eprint{1810.03626}.

\bibitem[{\citenamefont{Aristizabal~Sierra
  et~al.}(2018{\natexlab{a}})\citenamefont{Aristizabal~Sierra, De~Romeri, and
  Rojas}}]{AristizabalSierra:2018eqm}
\bibinfo{author}{\bibfnamefont{D.}~\bibnamefont{Aristizabal~Sierra}},
  \bibinfo{author}{\bibfnamefont{V.}~\bibnamefont{De~Romeri}},
  \bibnamefont{and} \bibinfo{author}{\bibfnamefont{N.}~\bibnamefont{Rojas}},
  \bibinfo{journal}{Phys. Rev.} \textbf{\bibinfo{volume}{D98}},
  \bibinfo{pages}{075018} (\bibinfo{year}{2018}{\natexlab{a}}),
  \eprint{1806.07424}.

\bibitem[{\citenamefont{Datta et~al.}(2019)\citenamefont{Datta, Dutta, Liao,
  Marfatia, and Strigari}}]{Datta:2018xty}
\bibinfo{author}{\bibfnamefont{A.}~\bibnamefont{Datta}},
  \bibinfo{author}{\bibfnamefont{B.}~\bibnamefont{Dutta}},
  \bibinfo{author}{\bibfnamefont{S.}~\bibnamefont{Liao}},
  \bibinfo{author}{\bibfnamefont{D.}~\bibnamefont{Marfatia}}, \bibnamefont{and}
  \bibinfo{author}{\bibfnamefont{L.~E.} \bibnamefont{Strigari}},
  \bibinfo{journal}{JHEP} \textbf{\bibinfo{volume}{01}}, \bibinfo{pages}{091}
  (\bibinfo{year}{2019}), \eprint{1808.02611}.

\bibitem[{\citenamefont{Ciuffoli et~al.}(2018)\citenamefont{Ciuffoli, Evslin,
  Fu, and Tang}}]{Ciuffoli:2018qem}
\bibinfo{author}{\bibfnamefont{E.}~\bibnamefont{Ciuffoli}},
  \bibinfo{author}{\bibfnamefont{J.}~\bibnamefont{Evslin}},
  \bibinfo{author}{\bibfnamefont{Q.}~\bibnamefont{Fu}}, \bibnamefont{and}
  \bibinfo{author}{\bibfnamefont{J.}~\bibnamefont{Tang}},
  \bibinfo{journal}{Phys. Rev.} \textbf{\bibinfo{volume}{D97}},
  \bibinfo{pages}{113003} (\bibinfo{year}{2018}), \eprint{1801.02166}.

\bibitem[{\citenamefont{Aristizabal~Sierra
  et~al.}(2019{\natexlab{a}})\citenamefont{Aristizabal~Sierra, Liao, and
  Marfatia}}]{AristizabalSierra:2019zmy}
\bibinfo{author}{\bibfnamefont{D.}~\bibnamefont{Aristizabal~Sierra}},
  \bibinfo{author}{\bibfnamefont{J.}~\bibnamefont{Liao}}, \bibnamefont{and}
  \bibinfo{author}{\bibfnamefont{D.}~\bibnamefont{Marfatia}},
  \bibinfo{journal}{JHEP} \textbf{\bibinfo{volume}{06}}, \bibinfo{pages}{141}
  (\bibinfo{year}{2019}{\natexlab{a}}), \eprint{1902.07398}.

\bibitem[{\citenamefont{Papoulias et~al.}(2020)\citenamefont{Papoulias, Kosmas,
  Sahu, Kota, and Hota}}]{Papoulias:2019lfi}
\bibinfo{author}{\bibfnamefont{D.~K.} \bibnamefont{Papoulias}},
  \bibinfo{author}{\bibfnamefont{T.~S.} \bibnamefont{Kosmas}},
  \bibinfo{author}{\bibfnamefont{R.}~\bibnamefont{Sahu}},
  \bibinfo{author}{\bibfnamefont{V.~K.~B.} \bibnamefont{Kota}},
  \bibnamefont{and} \bibinfo{author}{\bibfnamefont{M.}~\bibnamefont{Hota}},
  \bibinfo{journal}{Phys. Lett.} \textbf{\bibinfo{volume}{B800}},
  \bibinfo{pages}{135133} (\bibinfo{year}{2020}), \eprint{1903.03722}.

\bibitem[{\citenamefont{Kosmas et~al.}(2017)\citenamefont{Kosmas, Papoulias,
  Tortola, and Valle}}]{Kosmas:2017zbh}
\bibinfo{author}{\bibfnamefont{T.~S.} \bibnamefont{Kosmas}},
  \bibinfo{author}{\bibfnamefont{D.~K.} \bibnamefont{Papoulias}},
  \bibinfo{author}{\bibfnamefont{M.}~\bibnamefont{Tortola}}, \bibnamefont{and}
  \bibinfo{author}{\bibfnamefont{J.~W.~F.} \bibnamefont{Valle}},
  \bibinfo{journal}{Phys. Rev.} \textbf{\bibinfo{volume}{D96}},
  \bibinfo{pages}{063013} (\bibinfo{year}{2017}), \eprint{1703.00054}.

\bibitem[{\citenamefont{Blanco et~al.}(2019)\citenamefont{Blanco, Hooper, and
  Machado}}]{Blanco:2019vyp}
\bibinfo{author}{\bibfnamefont{C.}~\bibnamefont{Blanco}},
  \bibinfo{author}{\bibfnamefont{D.}~\bibnamefont{Hooper}}, \bibnamefont{and}
  \bibinfo{author}{\bibfnamefont{P.}~\bibnamefont{Machado}}
  (\bibinfo{year}{2019}), \eprint{1901.08094}.

\bibitem[{\citenamefont{Dutta et~al.}(2019{\natexlab{a}})\citenamefont{Dutta,
  Liao, Sinha, and Strigari}}]{Dutta:2019eml}
\bibinfo{author}{\bibfnamefont{B.}~\bibnamefont{Dutta}},
  \bibinfo{author}{\bibfnamefont{S.}~\bibnamefont{Liao}},
  \bibinfo{author}{\bibfnamefont{S.}~\bibnamefont{Sinha}}, \bibnamefont{and}
  \bibinfo{author}{\bibfnamefont{L.~E.} \bibnamefont{Strigari}},
  \bibinfo{journal}{Phys. Rev. Lett.} \textbf{\bibinfo{volume}{123}},
  \bibinfo{pages}{061801} (\bibinfo{year}{2019}{\natexlab{a}}),
  \eprint{1903.10666}.

\bibitem[{\citenamefont{Giunti}(2020)}]{Giunti:2019xpr}
\bibinfo{author}{\bibfnamefont{C.}~\bibnamefont{Giunti}},
  \bibinfo{journal}{Phys. Rev.} \textbf{\bibinfo{volume}{D101}},
  \bibinfo{pages}{035039} (\bibinfo{year}{2020}), \eprint{1909.00466}.

\bibitem[{\citenamefont{Dutta et~al.}(2019{\natexlab{b}})\citenamefont{Dutta,
  Kim, Liao, Park, Shin, and Strigari}}]{Dutta:2019nbn}
\bibinfo{author}{\bibfnamefont{B.}~\bibnamefont{Dutta}},
  \bibinfo{author}{\bibfnamefont{D.}~\bibnamefont{Kim}},
  \bibinfo{author}{\bibfnamefont{S.}~\bibnamefont{Liao}},
  \bibinfo{author}{\bibfnamefont{J.-C.} \bibnamefont{Park}},
  \bibinfo{author}{\bibfnamefont{S.}~\bibnamefont{Shin}}, \bibnamefont{and}
  \bibinfo{author}{\bibfnamefont{L.~E.} \bibnamefont{Strigari}}
  (\bibinfo{year}{2019}{\natexlab{b}}), \eprint{1906.10745}.

\bibitem[{\citenamefont{Mayet et~al.}(2016)}]{Mayet:2016zxu}
\bibinfo{author}{\bibfnamefont{F.}~\bibnamefont{Mayet}} \bibnamefont{et~al.},
  \bibinfo{journal}{Phys. Rept.} \textbf{\bibinfo{volume}{627}},
  \bibinfo{pages}{1} (\bibinfo{year}{2016}), \eprint{1602.03781}.

\bibitem[{\citenamefont{Battat et~al.}(2016)}]{Battat:2016pap}
\bibinfo{author}{\bibfnamefont{J.~B.~R.} \bibnamefont{Battat}}
  \bibnamefont{et~al.}, \bibinfo{journal}{Phys. Rept.}
  \textbf{\bibinfo{volume}{662}}, \bibinfo{pages}{1} (\bibinfo{year}{2016}),
  \eprint{1610.02396}.

\bibitem[{\citenamefont{Lewin and Smith}(1996)}]{Lewin:1995rx}
\bibinfo{author}{\bibfnamefont{J.~D.} \bibnamefont{Lewin}} \bibnamefont{and}
  \bibinfo{author}{\bibfnamefont{P.~F.} \bibnamefont{Smith}},
  \bibinfo{journal}{Astropart. Phys.} \textbf{\bibinfo{volume}{6}},
  \bibinfo{pages}{87} (\bibinfo{year}{1996}).

\bibitem[{\citenamefont{Angeli and Marinova}(2013)}]{Angeli:2013epw}
\bibinfo{author}{\bibfnamefont{I.}~\bibnamefont{Angeli}} \bibnamefont{and}
  \bibinfo{author}{\bibfnamefont{K.~P.} \bibnamefont{Marinova}},
  \bibinfo{journal}{Atom. Data Nucl. Data Tabl.} \textbf{\bibinfo{volume}{99}},
  \bibinfo{pages}{69} (\bibinfo{year}{2013}).

\bibitem[{\citenamefont{Freedman}(1974)}]{Freedman:1973yd}
\bibinfo{author}{\bibfnamefont{D.~Z.} \bibnamefont{Freedman}},
  \bibinfo{journal}{Phys. Rev.} \textbf{\bibinfo{volume}{D9}},
  \bibinfo{pages}{1389} (\bibinfo{year}{1974}).

\bibitem[{\citenamefont{Freedman et~al.}(1977)\citenamefont{Freedman, Schramm,
  and Tubbs}}]{Freedman:1977xn}
\bibinfo{author}{\bibfnamefont{D.~Z.} \bibnamefont{Freedman}},
  \bibinfo{author}{\bibfnamefont{D.~N.} \bibnamefont{Schramm}},
  \bibnamefont{and} \bibinfo{author}{\bibfnamefont{D.~L.} \bibnamefont{Tubbs}},
  \bibinfo{journal}{Ann. Rev. Nucl. Part. Sci.} \textbf{\bibinfo{volume}{27}},
  \bibinfo{pages}{167} (\bibinfo{year}{1977}).

\bibitem[{\citenamefont{Patrignani et~al.}(2016)}]{Olive:2016xmw}
\bibinfo{author}{\bibfnamefont{C.}~\bibnamefont{Patrignani}}
  \bibnamefont{et~al.} (\bibinfo{collaboration}{Particle Data Group}),
  \bibinfo{journal}{Chin. Phys.} \textbf{\bibinfo{volume}{C40}},
  \bibinfo{pages}{100001} (\bibinfo{year}{2016}).

\bibitem[{\citenamefont{Gondolo}(2002)}]{Gondolo:2002np}
\bibinfo{author}{\bibfnamefont{P.}~\bibnamefont{Gondolo}},
  \bibinfo{journal}{Phys. Rev.} \textbf{\bibinfo{volume}{D66}},
  \bibinfo{pages}{103513} (\bibinfo{year}{2002}), \eprint{hep-ph/0209110}.

\bibitem[{\citenamefont{O'Hare et~al.}(2015)\citenamefont{O'Hare, Green,
  Billard, Figueroa-Feliciano, and Strigari}}]{OHare:2015utx}
\bibinfo{author}{\bibfnamefont{C.~A.~J.} \bibnamefont{O'Hare}},
  \bibinfo{author}{\bibfnamefont{A.~M.} \bibnamefont{Green}},
  \bibinfo{author}{\bibfnamefont{J.}~\bibnamefont{Billard}},
  \bibinfo{author}{\bibfnamefont{E.}~\bibnamefont{Figueroa-Feliciano}},
  \bibnamefont{and} \bibinfo{author}{\bibfnamefont{L.~E.}
  \bibnamefont{Strigari}}, \bibinfo{journal}{Phys. Rev.}
  \textbf{\bibinfo{volume}{D92}}, \bibinfo{pages}{063518}
  (\bibinfo{year}{2015}), \eprint{1505.08061}.

\bibitem[{\citenamefont{Kopeikin}(2012)}]{Kopeikin:2012zz}
\bibinfo{author}{\bibfnamefont{V.~I.} \bibnamefont{Kopeikin}},
  \bibinfo{journal}{Phys. Atom. Nucl.} \textbf{\bibinfo{volume}{75}},
  \bibinfo{pages}{143} (\bibinfo{year}{2012}), \bibinfo{note}{[Yad.
  Fiz.75N2,165(2012)]}.

\bibitem[{\citenamefont{Shoemaker}(2017)}]{Shoemaker:2017lzs}
\bibinfo{author}{\bibfnamefont{I.~M.} \bibnamefont{Shoemaker}},
  \bibinfo{journal}{Phys. Rev.} \textbf{\bibinfo{volume}{D95}},
  \bibinfo{pages}{115028} (\bibinfo{year}{2017}), \eprint{1703.05774}.

\bibitem[{\citenamefont{Dutta et~al.}(2017)\citenamefont{Dutta, Liao, Strigari,
  and Walker}}]{Dutta:2017nht}
\bibinfo{author}{\bibfnamefont{B.}~\bibnamefont{Dutta}},
  \bibinfo{author}{\bibfnamefont{S.}~\bibnamefont{Liao}},
  \bibinfo{author}{\bibfnamefont{L.~E.} \bibnamefont{Strigari}},
  \bibnamefont{and} \bibinfo{author}{\bibfnamefont{J.~W.}
  \bibnamefont{Walker}}, \bibinfo{journal}{Phys. Lett.}
  \textbf{\bibinfo{volume}{B773}}, \bibinfo{pages}{242} (\bibinfo{year}{2017}),
  \eprint{1705.00661}.

\bibitem[{\citenamefont{Aristizabal~Sierra
  et~al.}(2018{\natexlab{b}})\citenamefont{Aristizabal~Sierra, Rojas, and
  Tytgat}}]{AristizabalSierra:2017joc}
\bibinfo{author}{\bibfnamefont{D.}~\bibnamefont{Aristizabal~Sierra}},
  \bibinfo{author}{\bibfnamefont{N.}~\bibnamefont{Rojas}}, \bibnamefont{and}
  \bibinfo{author}{\bibfnamefont{M.~H.~G.} \bibnamefont{Tytgat}},
  \bibinfo{journal}{JHEP} \textbf{\bibinfo{volume}{03}}, \bibinfo{pages}{197}
  (\bibinfo{year}{2018}{\natexlab{b}}), \eprint{1712.09667}.

\bibitem[{\citenamefont{Aristizabal~Sierra
  et~al.}(2019{\natexlab{b}})\citenamefont{Aristizabal~Sierra, De~Romeri, and
  Rojas}}]{AristizabalSierra:2019ufd}
\bibinfo{author}{\bibfnamefont{D.}~\bibnamefont{Aristizabal~Sierra}},
  \bibinfo{author}{\bibfnamefont{V.}~\bibnamefont{De~Romeri}},
  \bibnamefont{and} \bibinfo{author}{\bibfnamefont{N.}~\bibnamefont{Rojas}},
  \bibinfo{journal}{JHEP} \textbf{\bibinfo{volume}{09}}, \bibinfo{pages}{069}
  (\bibinfo{year}{2019}{\natexlab{b}}), \eprint{1906.01156}.

\bibitem[{\citenamefont{Miranda et~al.}(2020)\citenamefont{Miranda, Papoulias,
  T\'ortola, and Valle}}]{Miranda:2020zji}
\bibinfo{author}{\bibfnamefont{O.~G.} \bibnamefont{Miranda}},
  \bibinfo{author}{\bibfnamefont{D.~K.} \bibnamefont{Papoulias}},
  \bibinfo{author}{\bibfnamefont{M.}~\bibnamefont{T\'ortola}},
  \bibnamefont{and} \bibinfo{author}{\bibfnamefont{J.~W.~F.}
  \bibnamefont{Valle}} (\bibinfo{year}{2020}), \eprint{2002.01482}.

\bibitem[{\citenamefont{He et~al.}(1991{\natexlab{a}})\citenamefont{He, Joshi,
  Lew, and Volkas}}]{He:1990pn}
\bibinfo{author}{\bibfnamefont{X.~G.} \bibnamefont{He}},
  \bibinfo{author}{\bibfnamefont{G.~C.} \bibnamefont{Joshi}},
  \bibinfo{author}{\bibfnamefont{H.}~\bibnamefont{Lew}}, \bibnamefont{and}
  \bibinfo{author}{\bibfnamefont{R.~R.} \bibnamefont{Volkas}},
  \bibinfo{journal}{Phys. Rev.} \textbf{\bibinfo{volume}{D43}},
  \bibinfo{pages}{22} (\bibinfo{year}{1991}{\natexlab{a}}).

\bibitem[{\citenamefont{He et~al.}(1991{\natexlab{b}})\citenamefont{He, Joshi,
  Lew, and Volkas}}]{He:1991qd}
\bibinfo{author}{\bibfnamefont{X.-G.} \bibnamefont{He}},
  \bibinfo{author}{\bibfnamefont{G.~C.} \bibnamefont{Joshi}},
  \bibinfo{author}{\bibfnamefont{H.}~\bibnamefont{Lew}}, \bibnamefont{and}
  \bibinfo{author}{\bibfnamefont{R.~R.} \bibnamefont{Volkas}},
  \bibinfo{journal}{Phys. Rev.} \textbf{\bibinfo{volume}{D44}},
  \bibinfo{pages}{2118} (\bibinfo{year}{1991}{\natexlab{b}}).

\bibitem[{\citenamefont{Heeck}(2014)}]{Heeck:2014zfa}
\bibinfo{author}{\bibfnamefont{J.}~\bibnamefont{Heeck}},
  \bibinfo{journal}{Phys. Lett.} \textbf{\bibinfo{volume}{B739}},
  \bibinfo{pages}{256} (\bibinfo{year}{2014}), \eprint{1408.6845}.

\bibitem[{\citenamefont{Jeong et~al.}(2016)\citenamefont{Jeong, Kim, and
  Lee}}]{Jeong:2015bbi}
\bibinfo{author}{\bibfnamefont{Y.~S.} \bibnamefont{Jeong}},
  \bibinfo{author}{\bibfnamefont{C.~S.} \bibnamefont{Kim}}, \bibnamefont{and}
  \bibinfo{author}{\bibfnamefont{H.-S.} \bibnamefont{Lee}},
  \bibinfo{journal}{Int. J. Mod. Phys.} \textbf{\bibinfo{volume}{A31}},
  \bibinfo{pages}{1650059} (\bibinfo{year}{2016}), \eprint{1512.03179}.

\bibitem[{\citenamefont{Babu et~al.}(2017)\citenamefont{Babu, Friedland,
  Machado, and Mocioiu}}]{Babu:2017olk}
\bibinfo{author}{\bibfnamefont{K.~S.} \bibnamefont{Babu}},
  \bibinfo{author}{\bibfnamefont{A.}~\bibnamefont{Friedland}},
  \bibinfo{author}{\bibfnamefont{P.~A.~N.} \bibnamefont{Machado}},
  \bibnamefont{and} \bibinfo{author}{\bibfnamefont{I.}~\bibnamefont{Mocioiu}}
  (\bibinfo{year}{2017}), \eprint{1705.01822}.

\bibitem[{\citenamefont{Dutta et~al.}(2020)\citenamefont{Dutta, Ghosh, and
  Kumar}}]{Dutta:2020jsy}
\bibinfo{author}{\bibfnamefont{B.}~\bibnamefont{Dutta}},
  \bibinfo{author}{\bibfnamefont{S.}~\bibnamefont{Ghosh}}, \bibnamefont{and}
  \bibinfo{author}{\bibfnamefont{J.}~\bibnamefont{Kumar}}
  (\bibinfo{year}{2020}), \eprint{2002.01137}.

\bibitem[{\citenamefont{Dutta et~al.}(2019{\natexlab{c}})\citenamefont{Dutta,
  Ghosh, and Kumar}}]{Dutta:2019fxn}
\bibinfo{author}{\bibfnamefont{B.}~\bibnamefont{Dutta}},
  \bibinfo{author}{\bibfnamefont{S.}~\bibnamefont{Ghosh}}, \bibnamefont{and}
  \bibinfo{author}{\bibfnamefont{J.}~\bibnamefont{Kumar}},
  \bibinfo{journal}{Phys. Rev.} \textbf{\bibinfo{volume}{D100}},
  \bibinfo{pages}{075028} (\bibinfo{year}{2019}{\natexlab{c}}),
  \eprint{1905.02692}.

\bibitem[{\citenamefont{Farzan}(2015)}]{Farzan:2015doa}
\bibinfo{author}{\bibfnamefont{Y.}~\bibnamefont{Farzan}},
  \bibinfo{journal}{Phys. Lett.} \textbf{\bibinfo{volume}{B748}},
  \bibinfo{pages}{311} (\bibinfo{year}{2015}), \eprint{1505.06906}.

\bibitem[{\citenamefont{Farzan and Shoemaker}(2016)}]{Farzan:2015hkd}
\bibinfo{author}{\bibfnamefont{Y.}~\bibnamefont{Farzan}} \bibnamefont{and}
  \bibinfo{author}{\bibfnamefont{I.~M.} \bibnamefont{Shoemaker}},
  \bibinfo{journal}{JHEP} \textbf{\bibinfo{volume}{07}}, \bibinfo{pages}{033}
  (\bibinfo{year}{2016}), \eprint{1512.09147}.

\bibitem[{\citenamefont{Aristizabal~Sierra
  et~al.}(2019{\natexlab{c}})\citenamefont{Aristizabal~Sierra, Dutta, Liao, and
  Strigari}}]{AristizabalSierra:2019ykk}
\bibinfo{author}{\bibfnamefont{D.}~\bibnamefont{Aristizabal~Sierra}},
  \bibinfo{author}{\bibfnamefont{B.}~\bibnamefont{Dutta}},
  \bibinfo{author}{\bibfnamefont{S.}~\bibnamefont{Liao}}, \bibnamefont{and}
  \bibinfo{author}{\bibfnamefont{L.~E.} \bibnamefont{Strigari}},
  \bibinfo{journal}{JHEP} \textbf{\bibinfo{volume}{12}}, \bibinfo{pages}{124}
  (\bibinfo{year}{2019}{\natexlab{c}}), \eprint{1910.12437}.

\bibitem[{\citenamefont{Crivellin et~al.}(2014)\citenamefont{Crivellin,
  Hoferichter, and Procura}}]{Crivellin:2013ipa}
\bibinfo{author}{\bibfnamefont{A.}~\bibnamefont{Crivellin}},
  \bibinfo{author}{\bibfnamefont{M.}~\bibnamefont{Hoferichter}},
  \bibnamefont{and} \bibinfo{author}{\bibfnamefont{M.}~\bibnamefont{Procura}},
  \bibinfo{journal}{Phys. Rev.} \textbf{\bibinfo{volume}{D89}},
  \bibinfo{pages}{054021} (\bibinfo{year}{2014}), \eprint{1312.4951}.

\bibitem[{\citenamefont{Hoferichter et~al.}(2015)\citenamefont{Hoferichter,
  Ruiz~de Elvira, Kubis, and Meißner}}]{Hoferichter:2015dsa}
\bibinfo{author}{\bibfnamefont{M.}~\bibnamefont{Hoferichter}},
  \bibinfo{author}{\bibfnamefont{J.}~\bibnamefont{Ruiz~de Elvira}},
  \bibinfo{author}{\bibfnamefont{B.}~\bibnamefont{Kubis}}, \bibnamefont{and}
  \bibinfo{author}{\bibfnamefont{U.-G.} \bibnamefont{Meißner}},
  \bibinfo{journal}{Phys. Rev. Lett.} \textbf{\bibinfo{volume}{115}},
  \bibinfo{pages}{092301} (\bibinfo{year}{2015}), \eprint{1506.04142}.

\bibitem[{\citenamefont{Ellis et~al.}(2000)\citenamefont{Ellis, Ferstl, and
  Olive}}]{Ellis:2000ds}
\bibinfo{author}{\bibfnamefont{J.~R.} \bibnamefont{Ellis}},
  \bibinfo{author}{\bibfnamefont{A.}~\bibnamefont{Ferstl}}, \bibnamefont{and}
  \bibinfo{author}{\bibfnamefont{K.~A.} \bibnamefont{Olive}},
  \bibinfo{journal}{Phys. Lett.} \textbf{\bibinfo{volume}{B481}},
  \bibinfo{pages}{304} (\bibinfo{year}{2000}), \eprint{hep-ph/0001005}.

\bibitem[{\citenamefont{Nelson and Walsh}(2008{\natexlab{a}})}]{Nelson:2007yq}
\bibinfo{author}{\bibfnamefont{A.~E.} \bibnamefont{Nelson}} \bibnamefont{and}
  \bibinfo{author}{\bibfnamefont{J.}~\bibnamefont{Walsh}},
  \bibinfo{journal}{Phys. Rev.} \textbf{\bibinfo{volume}{D77}},
  \bibinfo{pages}{033001} (\bibinfo{year}{2008}{\natexlab{a}}),
  \eprint{0711.1363}.

\bibitem[{\citenamefont{Nelson and Walsh}(2008{\natexlab{b}})}]{Nelson:2008tn}
\bibinfo{author}{\bibfnamefont{A.~E.} \bibnamefont{Nelson}} \bibnamefont{and}
  \bibinfo{author}{\bibfnamefont{J.}~\bibnamefont{Walsh}},
  \bibinfo{journal}{Phys. Rev.} \textbf{\bibinfo{volume}{D77}},
  \bibinfo{pages}{095006} (\bibinfo{year}{2008}{\natexlab{b}}),
  \eprint{0802.0762}.

\bibitem[{\citenamefont{Aghanim et~al.}(2018)}]{Aghanim:2018eyx}
\bibinfo{author}{\bibfnamefont{N.}~\bibnamefont{Aghanim}} \bibnamefont{et~al.}
  (\bibinfo{collaboration}{Planck}) (\bibinfo{year}{2018}),
  \eprint{1807.06209}.

\bibitem[{\citenamefont{Kamada and Yu}(2015)}]{Kamada:2015era}
\bibinfo{author}{\bibfnamefont{A.}~\bibnamefont{Kamada}} \bibnamefont{and}
  \bibinfo{author}{\bibfnamefont{H.-B.} \bibnamefont{Yu}},
  \bibinfo{journal}{Phys. Rev.} \textbf{\bibinfo{volume}{D92}},
  \bibinfo{pages}{113004} (\bibinfo{year}{2015}), \eprint{1504.00711}.

\bibitem[{\citenamefont{Knapen et~al.}(2017)\citenamefont{Knapen, Lin, and
  Zurek}}]{Knapen:2017xzo}
\bibinfo{author}{\bibfnamefont{S.}~\bibnamefont{Knapen}},
  \bibinfo{author}{\bibfnamefont{T.}~\bibnamefont{Lin}}, \bibnamefont{and}
  \bibinfo{author}{\bibfnamefont{K.~M.} \bibnamefont{Zurek}},
  \bibinfo{journal}{Phys. Rev.} \textbf{\bibinfo{volume}{D96}},
  \bibinfo{pages}{115021} (\bibinfo{year}{2017}), \eprint{1709.07882}.

\bibitem[{\citenamefont{Escudero et~al.}(2019)\citenamefont{Escudero, Hooper,
  Krnjaic, and Pierre}}]{Escudero:2019gzq}
\bibinfo{author}{\bibfnamefont{M.}~\bibnamefont{Escudero}},
  \bibinfo{author}{\bibfnamefont{D.}~\bibnamefont{Hooper}},
  \bibinfo{author}{\bibfnamefont{G.}~\bibnamefont{Krnjaic}}, \bibnamefont{and}
  \bibinfo{author}{\bibfnamefont{M.}~\bibnamefont{Pierre}},
  \bibinfo{journal}{JHEP} \textbf{\bibinfo{volume}{03}}, \bibinfo{pages}{071}
  (\bibinfo{year}{2019}), \eprint{1901.02010}.

\bibitem[{\citenamefont{Sabti et~al.}(2020)\citenamefont{Sabti, Alvey,
  Escudero, Fairbairn, and Blas}}]{Sabti:2019mhn}
\bibinfo{author}{\bibfnamefont{N.}~\bibnamefont{Sabti}},
  \bibinfo{author}{\bibfnamefont{J.}~\bibnamefont{Alvey}},
  \bibinfo{author}{\bibfnamefont{M.}~\bibnamefont{Escudero}},
  \bibinfo{author}{\bibfnamefont{M.}~\bibnamefont{Fairbairn}},
  \bibnamefont{and} \bibinfo{author}{\bibfnamefont{D.}~\bibnamefont{Blas}},
  \bibinfo{journal}{JCAP} \textbf{\bibinfo{volume}{2001}}, \bibinfo{pages}{004}
  (\bibinfo{year}{2020}), \eprint{1910.01649}.

\bibitem[{\citenamefont{Akimov et~al.}(2019)}]{Akimov:2019xdj}
\bibinfo{author}{\bibfnamefont{D.}~\bibnamefont{Akimov}} \bibnamefont{et~al.}
  (\bibinfo{collaboration}{COHERENT}) (\bibinfo{year}{2019}),
  \eprint{1911.06422}.

\bibitem[{\citenamefont{Ibe et~al.}(2018)\citenamefont{Ibe, Nakano, Shoji, and
  Suzuki}}]{Ibe:2017yqa}
\bibinfo{author}{\bibfnamefont{M.}~\bibnamefont{Ibe}},
  \bibinfo{author}{\bibfnamefont{W.}~\bibnamefont{Nakano}},
  \bibinfo{author}{\bibfnamefont{Y.}~\bibnamefont{Shoji}}, \bibnamefont{and}
  \bibinfo{author}{\bibfnamefont{K.}~\bibnamefont{Suzuki}},
  \bibinfo{journal}{JHEP} \textbf{\bibinfo{volume}{03}}, \bibinfo{pages}{194}
  (\bibinfo{year}{2018}), \eprint{1707.07258}.

\bibitem[{\citenamefont{Aprile et~al.}(2019)}]{Aprile:2019jmx}
\bibinfo{author}{\bibfnamefont{E.}~\bibnamefont{Aprile}} \bibnamefont{et~al.}
  (\bibinfo{collaboration}{XENON}), \bibinfo{journal}{Phys. Rev. Lett.}
  \textbf{\bibinfo{volume}{123}}, \bibinfo{pages}{241803}
  (\bibinfo{year}{2019}), \eprint{1907.12771}.

\end{thebibliography}
\end{document}